\documentclass[journal, 10pt]{IEEEtran}

\usepackage{multirow}
\usepackage[font=scriptsize,caption=false,labelsep=space]{subfig}
\usepackage{mathrsfs}
\usepackage{graphicx,float,wrapfig,epstopdf,amsmath}
\usepackage[]{algorithmicx}
\usepackage{algpseudocode,algorithm}
\usepackage{balance}
\usepackage{enumitem}
\usepackage{cuted}

\usepackage{mathtools,lipsum}
\usepackage{amsmath}
\usepackage{amssymb}
\usepackage{amsthm}
\usepackage{graphicx}
\usepackage{epstopdf}
\usepackage{times}
\usepackage{textcomp,cite}

\usepackage{url}
\usepackage{hyperref}

\usepackage{multirow}
\usepackage{threeparttable}
\graphicspath{{./figures/}}

\DeclareMathOperator*{\argmax}{argmax} 
\DeclareMathOperator*{\minimize}{minimize} 
\DeclareMathOperator*{\subjectto}{subject \hspace{3pt} to:\hspace{3pt}} 

\usepackage[table]{xcolor}
\definecolor{infocolor}{RGB}{213,229,255}
\definecolor{inteins}{RGB}{128,179,255}
\definecolor{color1}{RGB}{199,209,232}
\definecolor{color2}{RGB}{230,231,233}

\usepackage[table]{xcolor}
\newtheorem{theorem}{Theorem}

\newtheorem{lemma}[theorem]{Lemma}

\usepackage{enumitem}

\begin{document}

	\title{Terahertz-Band Channel and Beam Split Estimation via Array Perturbation Model
	}

	\author{\IEEEauthorblockN{Ahmet M. Elbir, \textit{Senior Member, IEEE,} Wei Shi,
			Anastasios K. Papazafeiropoulos, \textit{Senior Member, IEEE,}
			Pandelis Kourtessis,
			and  Symeon Chatzinotas, \textit{Fellow, IEEE}
		}
		\thanks{This work was supported in part by the Horizon Project TERRAMETA and the Natural Sciences and Engineering Research Council of Canada	 (NSERC) and Ericsson Canada.}
		\thanks{A. M. Elbir is with 
			Interdisciplinary Centre for Security, Reliability and Trust (SnT) at the University of Luxembourg, Luxembourg (e-mail: ahmetmelbir@ieee.org).} 
		\thanks{W. Shi is with the	School of Information Technology, Carleton
			University, Ottawa, Canada (e-mail: wei.shi@carleton.ca).}
		\thanks{A. K. Papazafeiropoulos is with the CIS Research Group, University of Hertfordshire, Hatfield, U. K., and SnT at the University of Luxembourg, Luxembourg (e-mail: tapapazaf@gmail.com). }
		\thanks{P. Kourtessis is with the CIS Research Group, University of Hertfordshire, Hatfield, U. K. (e-mail: p.kourtessis@herts.ac.uk).}
		
		\thanks{S. Chatzinotas is with the SnT at the University of Luxembourg, Luxembourg (e-mail: symeon.chatzinotas@uni.lu). }
		
	}

		\maketitle
	\begin{abstract}
		For the demonstration of ultra-wideband bandwidth and pencil-beamforming, the terahertz (THz)-band has been envisioned as one of the key enabling technologies for the sixth generation networks. However, the acquisition of the THz channel entails several unique challenges such as severe path loss and beam-split. Prior works usually employ ultra-massive arrays and additional hardware components comprised of time-delayers to compensate for these loses. In order to provide a cost-effective solution, this paper introduces a sparse-Bayesian-learning (SBL) technique for joint channel and beam-split estimation. Specifically, we first model the beam-split as an array perturbation inspired from array signal processing. Next, a low-complexity approach is developed by exploiting the line-of-sight-dominant feature of THz channel to reduce the computational complexity involved in the proposed SBL technique for channel estimation (SBCE). Additionally, based on federated-learning, we implement a model-free technique to the proposed model-based SBCE solution. Further to that, we examine the near-field considerations of THz channel, and introduce the range-dependent near-field beam-split. The theoretical performance bounds, i.e., Cram\'er-Rao lower bounds, are derived both for near- and far-field parameters, e.g., user directions, beam-split and ranges. Numerical simulations demonstrate that SBCE outperforms the existing approaches and exhibits lower hardware cost.
		
	\end{abstract}
	
	
	
	\maketitle
	
	\section{Introduction}
	
	The definition of the THz band varies among different IEEE communities. While IEEE Terahertz Technology and Applications Committee focus on $0.3-3$ THz band, the IEEE Transactions on Terahertz Science and Technology Journal studies $0.3-10$ THz~\cite{ummimoTareqOverview,elbir2021JointRadarComm}. On the other hand, $0.1-10$ THz band has been used for THz band in the recent studies on wireless communications including a large overlap with millimeter wave (mmWave), e.g., $0.03-0.3$ THz~\cite{ummimoTareqOverview}.	Furthermore,  the Federal Communication Commission (FCC) has designated the frequency bands of $116-123$ GHz, $174.8-182$ GHz, $185-190$ GHz and $244-246$ GHz for the unlicensed use for the future spectrum horizon~\cite{thz_bands_BibEntry2019Jun}.

\begin{figure*}[t]
		\centering
		\subfloat[]{\includegraphics[draft=false,width=.33\textwidth]{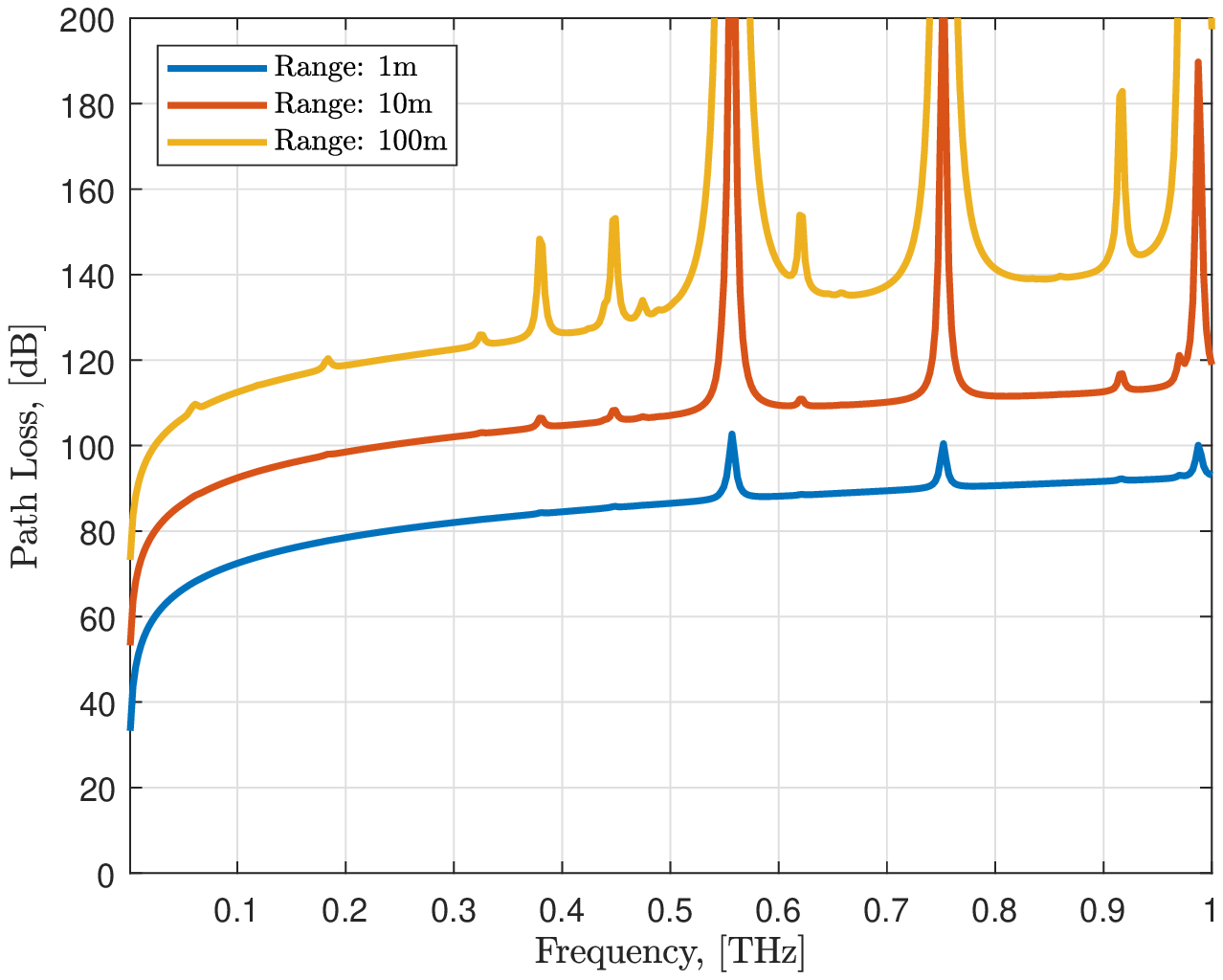} } 
		\subfloat[]{\includegraphics[draft=false,width=.33\textwidth]{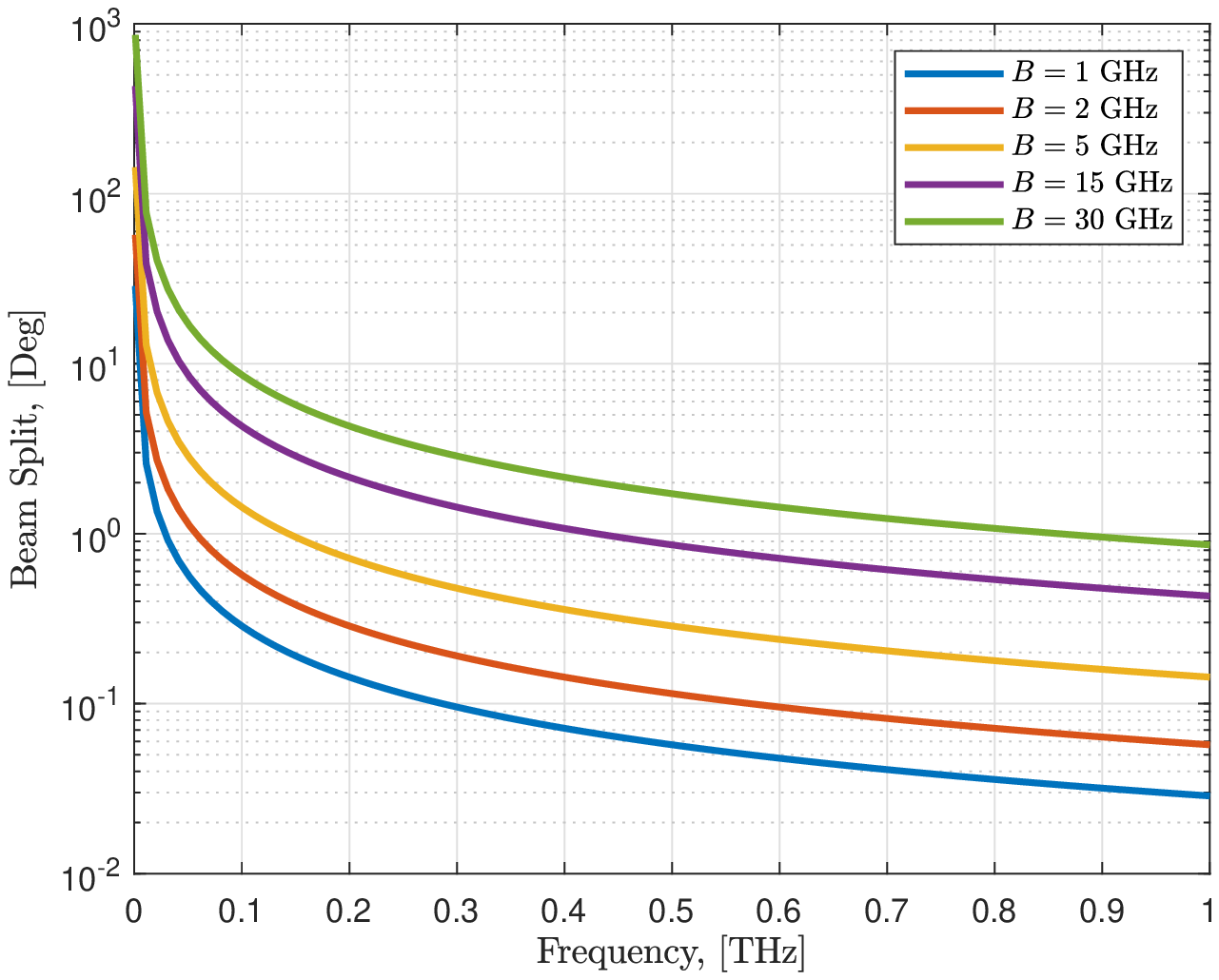} }
		\subfloat[]{\includegraphics[draft=false,width=.33\textwidth]{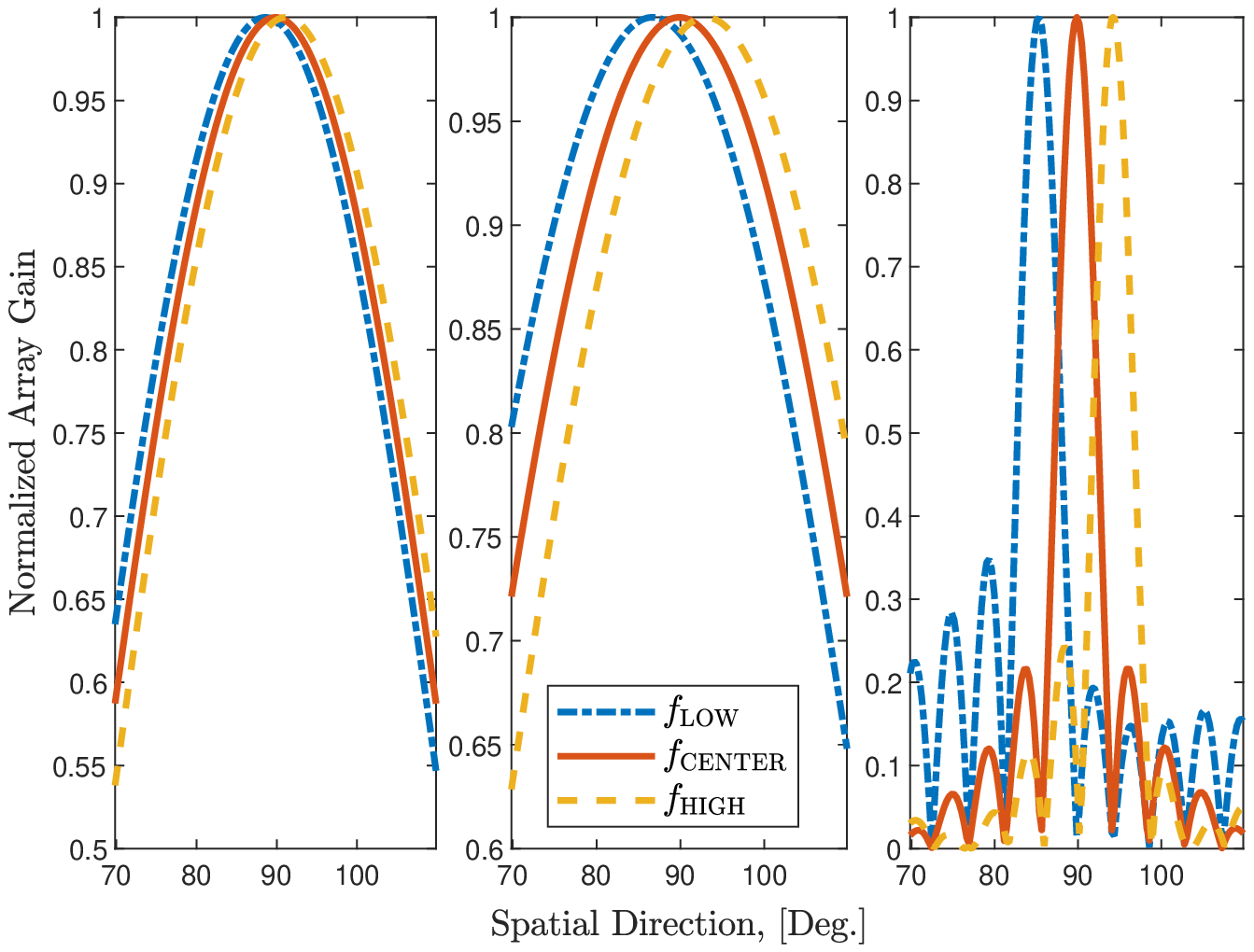} } 
		\caption{THz-band characteristics: (a) path loss (in dB) due to molecular absorption for various transmission ranges,  (b) beam-split (in degrees) \textcolor{black}{with respect to the carrier frequency $f_c \in [50, 1000]$ GHz} for different bandwidths, and  (c) normalized array gain with respect to spatial direction at low, center and high end subcarriers for (Left) $3.5$ GHz, $B=0.1$ GHz; (Middle) $28$ GHz, $B=2$ GHz; and (Right) $300$ GHz, $30$ GHz, respectively.    	}
		\label{fig_Diag}
	\end{figure*}

	Compared to its mmWave counterpart, the THz-band channel exhibits several THz-specific challenges that should be taken into account. These unique THz features, among others, include high path loss due to spreading loss and molecular absorption, shorter transmission rage, distance-dependent bandwidth and beam-split (see, e.g., Fig.~\ref{fig_Diag})~\cite{ummimoTareqOverview,thz_Akyildiz2022May,elbir2022Aug_THz_ISAC}.
	Furthermore, the THz channels are extremely-sparse. Hence, they are usually characterized as line-of-sight (LoS)-dominant and non-LoS (NLoS)-assisting models~\cite{ummimoTareqOverview,ummimoHBThzSVModel,thz_beamSplit,teraMIMO,ummimoTareq,faisal2020ultramassive,Ning2021Jan}. In comparison, the mmWave channel involves both LoS and NLoS paths, which are significant while LoS path is dominant in THz model~\cite{beamSquint_FeiFei_Wang2019Oct}. In addition, to compensate the severe path loss, analogous to massive multiple-input multiple-output (MIMO) array in mmWave design, ultra-massive (UM) MIMO array configurations are proposed for THz~\cite{ummimoComMagYeLi,elbir2021JointRadarComm}. The UM-MIMO design comprises huge number of antennas which are densely-positioned (e.g., $5\times 5 $ $\text{cm}^2$) due to extremely short wavelength~\cite{elbir2022Aug_THz_ISAC}. As a result, the usage of dedicated radio-frequency (RF) chain for each antenna element involves extreme hardware cost and labor. The hybrid design configurations, i.e., joint usage of analog and digital beamformers seems to be major possible solution as it was first envisioned for mmWave massive MIMO systems in 5G~\cite{mimoRHeath,mimoOverview,mimoChannelModel1,mimoHybridLeus3,elbir2022Nov_SPM_beamforming}. In order to exploit low cost system design, the hybrid architecture involve a few (large) number of digital (analog) beamformers. Although the digital beamformers are subcarrier-dependent (SD), the analog phase shifters are designed as subcarrier-independent (SI) components. This causes \textit{beam-split} phenomenon, i.e., the beams generated by the analog beamformer according to a single frequency split  at different subcarriers and they look at different directions (Fig.~\ref{fig_Diag}) because of ultra-wide bandwidth and large number of antennas~\cite{thz_beamSplit,thz_channelEst_beamsplitPatternDetection_L_Dai}. In mmWave, at which the subcarrier frequencies are relatively closer than THz, \textit{beam-squint} is the term broadly used to describe the same phenomenon~\cite{beamSquint_FeiFei_Wang2019Oct,thz_doa_est_beamSplit_Wu2019Aug,beamSquintWang2019Nov}. The beam-split affects the system performance and causes severe degradations in terms of spectral efficiency, normalized mean-squared-error (NMSE) and bit-error-rate (BER). For instance, the beam-split is approximately $4^\circ $ ($0.4^\circ$) for $0.3$ THz with $30$ GHz ($60$ GHz with $1$ GHz) bandwidth, respectively for a broadside user~\cite{elbir2022Aug_THz_ISAC}. \textcolor{black}{Fig.~\ref{fig_Diag}b shows beam-split (in degrees) with respect to the carrier frequency for different bandwidths. Large beam-split occurs at lower  frequencies since the bandwidth is relatively larger with respect to the carrier frequency, and it becomes smaller as the  carrier frequency increases.       }

	With the aforementioned THz-specific features, THz channel estimation is regarded as an even more challenging problem than that of mmWave. In prior studies, THz channel estimation has been investigated in~\cite{channelEstThz,thz_SBL_no_BS_Srivastava2021Sep,channelEstThz4,thz_CE_GAN_Balevi2021Jan,thz_CE_singleCarrier_NB_Peng2019Jan,thzCE_kernel,thzCE_CNN,dovelos_THz_CE_channelEstThz2,thz_channelEst_beamsplitPatternDetection_L_Dai,elbir2022Jul_THz_CE_FL}. However, most of these works either ignore the effect of beam-split~\cite{channelEstThz,thz_SBL_no_BS_Srivastava2021Sep,channelEstThz4,thz_CE_GAN_Balevi2021Jan,thz_CE_RF_impair_Sha2021Apr} or consider only the narrowband case~\cite{thz_CE_singleCarrier_NB_Peng2019Jan,thzCE_kernel,thzCE_CNN}, without exploiting the ultra-wide bandwidth property, the key reason of climbing to THz-band. Further to that, an  machine learning (ML)-based SBL architecture used in~\cite{thz_SBL_Geoffrey_Gao2022May} for mmWave channel estimation in the presence of beam-squint assumes SI steering vectors to construct the wideband channel. Hence, its performance is limited and, indeed, no performance improvement was observed by enhancing the grid resolution. In fact, the beam-split compensation requires certain signal processing or hardware techniques to be handled properly.
	
	Despite the prominence of THz channel estimation, in the literature, there are only a few recent works on beam-split mitigation. The existing solutions are categorized into two classes, i.e., hardware-based techniques~\cite{dovelos_THz_CE_channelEstThz2} and algorithmic methods~\cite{thz_channelEst_beamsplitPatternDetection_L_Dai,elbir2022Jul_THz_CE_FL}. The first category of solutions consider employing time delayer (TD) networks together with phase shifters to realize virtual SD analog beamformers to mitigate beam-split. In particular, ~\cite{dovelos_THz_CE_channelEstThz2} devises a generalized simultaneous orthogonal matching pursuit (GSOMP) technique by exploiting the SD information collected via TD network hence achieves close to minimum MSE (MSE) performance. However, these solutions require additional hardware, i.e., each phase shifter is connected to multiple TDs, each of which consumes approximately $100$ mW, which is more than that of a phase shifter ($40$ mW) in THz~\cite{elbir2022Aug_THz_ISAC}. The second category of solutions do not employ additional hardware components. Instead, advanced signal processing techniques have been proposed to compensate beam-split. Specifically, an OMP-based beam-split pattern detection (BSPD) approach was proposed in~\cite{thz_channelEst_beamsplitPatternDetection_L_Dai} for the recovery of the support pattern  among all subcarriers in the beamspace and construct one-to-one match between the physical and spatial (i.e., deviated due to beam-split in the beamspace) directions. Also,  \cite{spatialWidebandWang2018May} proposed an angular-delay rotation method, which suffers from coarse beam-split estimation and high training overhead due to the use of complete discrete Fourier transform (DFT) matrix. In \cite{elbir2022Jul_THz_CE_FL},  a beamspace support alignment (BSA) technique was introduced to align the deviated spatial beam directions among the subcarriers. Although both BSPD and BSA are based on OMP, the latter exhibits lower NMSE for THz channel estimation. Nevertheless, both methods suffer from inaccurate support detection and low precision for estimating the physical channel directions.
	
	Due to short transmission range in THz, near-field spherical-wave models are  considered for THz applications~\cite{thz_nearField_BeamSplit_Cui2021Sep,elbir2022Aug_THz_ISAC}. In far-field model, the transmitted signal arrives at the users as plane-wave whereas the plane wavefront is spherical in the near-field if the transmission range is shorter than the Fraunhofer distance~\cite{nearFieldDOA_Jiang2013Apr,thz_nearField_BeamSplit_Cui2021Sep}. Thus, the near-field effects should be taken into account for accurate channel modeling. In~\cite{thz_nearField_BeamSplit_Cui2021Sep}, THz near-field beamforming problem has been considered and TD network-based approach was proposed while the THz channel was assumed to be known a prior. In a different line of research, the beam-squint problem is investigated in~\cite{sbl_CE_beamSquint_Kim2021Feb} for the massive MIMO system with low resolution analog-digital converters. 
	
	In this paper, THz channel estimation in the presence of beam-split is examined. By exploiting the extreme sparsity of THz channel, we first approach the problem from the sparse recovery optimization perspective. Then, we introduce a novel approach to model the beam-split as an array perturbation as inspired by the array imperfection models, e.g., mutual coupling, gain-phase mismatch, in array signal processing context~\cite{elbir2016Jul_Sparse_MutualCoupling,sbl_doa_Liu2012Sep,sbl_CE_mmWave_Tang2018Jun}. The array perturbation model allows us to establish a linear transformation between the nominal (constructed via physical directions) and actual (beam-split corrupted) steering vectors. Next, we propose a sparse Bayesian learning (SBL) approach to jointly estimate both THz channel and beam-split. 	The SBL method has been shown to be effective  for sparse signal reconstruction from underdetermined observations~\cite{sbl1_tipping2001sparse,sbl2_Dempster1977Sep,sbl3_Radich1997Apr}. Compared to other existing signal estimation techniques, e.g., \textit{mu}ltiple \textit{si}gnal \textit{c}lassification (MUSIC)~\cite{music} and the $\ell_p$-norm ($0 \leq p \leq 1$) techniques such as compressed sensing (CS)~\cite{compressedSensing_SPM_Candes2008Mar}, SBL outperforms these techniques in terms of precision and convergence~\cite{sbl_versus_CS_Tan2009Jan}.
	Although SBL has been widely used for both channel estimation~\cite{sbl_CE_mmWave_Tang2018Jun} and direction-of-arrival (DoA) estimation~\cite{sbl_doa_Liu2012Sep,sbl_doa2_Liu2013May}, the proposed SBL-based channel estimation (SBCE) approach differentiates from the these studies by jointly estimating multiple hyperparameters, e.g., physical channel directions and beam-split as well as array the perturbation-based beam-split model. The proposed SBCE approach is advantageous in terms of complexity since it does not require any additional hardware components as in TD-based works, and it exhibits close to optimum performance for both THz channel and beam-split estimation. Furthermore, the estimation of beam-split allows one to design the hybrid beamformers in massive/UM MIMO systems more accurately in a simpler way.
	The main contributions of this work are summarized as follows:
	
	\begin{enumerate}
		\item We propose a novel approach to model beam-split as an array perturbation, and design a transformation matrix between the nominal and actual steering vectors in order to  estimate both  THz channel and beam-split.
		\item An SBL approach is devised to jointly estimate the physical channel directions and beam-split. In order to reduce the computational complexity involved in SBL iterations, we exploit the LoS-dominant feature of THz channel and design the array perturbation matrix based on a single unknown parameter, i.e., the beam-split. Thus, instead of designing the array perturbations via a full matrix with distinct elements, we consider a diagonal structure, that can be easily represented via only beam-split knowledge.
		\item We also propose a model-free approach based on federated learning (FL) to ease the communication overhead while maintaining satisfactory NMSE performance. Different from our preliminary work in~\cite{elbir2022Jul_THz_CE_FL}, we consider SBCE method for data labeling, hence it is called sparse Bayesian FL (SBFL).
		\item  We investigate the near-field considerations for THz transmission, and derive the near-field beam-split, which is both range- and direction-dependent.
		\item  In addition, the theoretical performance bounds for both physical channel directions and beam-split estimation are examined and the corresponding Cram\'er-Rao lower bounds (CRBs) have been derived. 
	\end{enumerate}

	\textit{Paper Organization:} The rest of the paper is organized as follows. In Sec.~\ref{sec:probForm}, we present the array signal model and formulate the THz channel estimation problem. Next, our proposed SBCE approach is introduced in Sec.~\ref{sec:SBCE}. The near-field model for beam-split and the proposed SBFL approach are given in Sec.~\ref{sec:nearField} and Sec.~\ref{sec:SBFL}, respectively. We present extensive numerical simulations in Sec.~\ref{sec:Sim}, and finalize the paper  with conclusions in Sec.~\ref{sec:Conc}.
	
	\textit{Notation:}  $(\cdot)^\textsf{T}$ and $(\cdot)^{\textsf{H}}$ to represent the transpose and conjugate-transpose operations, respectively. For a matrix $\mathbf{A}$; $[\mathbf{A}]_{ij}$ and $[\mathbf{A}]_k$  correspond to the $(i,j)$th entry and  $k$th column while $\mathbf{A}^{\dagger}$ denotes the Moore-Penrose pseudo-inverse of $\mathbf{A}$. A unit matrix of size $N$ is represented by $\mathbf{I}_N$, $\nabla$ represents the gradient operation, $\xi (a) = \frac{\sin N \pi a}{N \sin \pi a }$ is the Dirichlet sinc function, and $\mathrm{Tr}\{\cdot \}$ stands for the trace operation. $|| \cdot ||_0$, $|| \cdot ||_1$, $|| \cdot ||_2$ and $|| \cdot ||_\mathcal{F}$ denote the $\ell_0$, $\ell_1$, $\ell_2$ and Frobenius norms, respectively.

	\section{Signal Model}
	\label{sec:probForm}
	We consider a downlink scenario for  multi-user wideband MIMO system with OFDM (orthogonal frequency division multiplexing) employing $M$ subcarriers. We assume that the base station (BS) employs $N_\mathrm{T}$ antennas together with $N_\mathrm{RF}$ RF chains to communicate with $K$ single-antenna users. Hence, the data symbols, $\mathbf{s}[m] = [s_1[m],\cdots,s_K[m]]^\textsf{T}$ ($m\in \mathcal{M} = \{1,\cdots, M\}$) are precoded via the SD baseband beamformer $\mathbf{F}_\mathrm{BB}[m]\in \mathbb{C}^{K\times K}$. The resulting  signal $\mathbf{F}_\mathrm{BB}[m]\mathbf{s}[m]\in \mathbb{C}^{K}$ is constructed in the time domain by employing an $M$-point inverse DFT (IDFT). Then, the SI analog beamformers $\mathbf{F}_\mathrm{RF}\in \mathbb{C}^{N_\mathrm{T}\times N_\mathrm{RF}}$ ($N_\mathrm{RF}=K<{N}_\mathrm{T}$) are used to process the cyclic-prefix (CP)-added signal. The analog beamformers, which  are SI and realized with phase-shifters, have constant-modulus constraint, i.e., $|[\mathbf{F}_\mathrm{RF}]_{i,j}| = \frac{1}{\sqrt{N_\mathrm{T}}}$ for $i = 1,\cdots, N_\mathrm{RF}$, $j = 1,\cdots, N_\mathrm{T}$, and the constraint $\sum_{m = 1}^{M}||\mathbf{F}_\mathrm{RF}\mathbf{F}_\mathrm{BB}[m]  ||_\mathcal{F}^2 = MK$ enforces the transmission power.  We can define the $N_\mathrm{T}\times 1$ transmitted signal as
	\begin{align}
	\mathbf{t}[m] = \mathbf{F}_\mathrm{RF}\mathbf{F}_\mathrm{BB}[m]\mathbf{s}[m],
	\end{align}
	which is then received at the $k$th user as
	\begin{align}
	\label{receivedSignal}
	{y}_{k}[m] = \mathbf{h}_{k}^\textsf{T}[m]\mathbf{t}[m]  + {w}_k[m],
	\end{align}
	where ${w}_k[m]\in \mathbb{C}$ corresponds to the additive white Gaussian noise (AWGN), and we have ${w}_k[m] \sim \mathcal{CN}({0},\mu^2)$.

	\subsection{THz Channel Model}
	The THz channel is modeled by combination of a single LoS path together with a few NLoS paths, which are weak due to reflection, scattering and refraction~\cite{ummimoTareqOverview,ummimoHBThzSVModel,thz_beamSplit,teraMIMO}. In addition, the ray-tracing techniques for THz communication also assume the extreme sparsity of the THz channel and state that the channel is dominated by the LoS component for the graphene-based nano-transceivers \cite{ummimoTareqOverview} while the other channel models, e.g., 3GPP channel model \cite{3gppChannelModelThz,umMIMO3gpp_IRS} are also widely used for THz transmission.	Then,  the THz delay-$d$ MIMO channel is given in time domain as
	\begin{align}
	\label{channelTimeDomain}
	\bar{\mathbf{h}}_k(d) = &\sqrt{\frac{N_\mathrm{T}}{L}}    \bigg(  \underbrace{p(d T_s - \tau_{k,1})\alpha_{k}^{1} \mathbf{a}(\vartheta_{k,1}) }_{\text{LoS}} 
	\nonumber \\
	&
	\hspace{20pt}
	+  \underbrace{\sum_{l =2}^{L} p(d T_s - \tau_{k,l}) \alpha_{k}^{l} \mathbf{a}(\vartheta_{k,l})}_{\text{NLoS}}   \bigg),
	\end{align}
	where $\alpha_{k}^{l}\in\mathbb{C}$ denotes the complex path gain and  $L$ is the total number of paths. The time delay of the $l$th path is denoted by $\tau_{k,l}$, for which $l=1$ corresponds to the LoS path, and $p(\cdot)$ denotes the pulse-shaping function for $T_s$-signaling.  Also, ${\mathbf{a}}(\vartheta_{k,l}) \in\mathbb{C}^{N_\mathrm{T}}  $ denotes the array steering vector corresponding to the physical direction for the $l$th path of the $k$th user  $\vartheta_{k,l} = \sin \tilde{\vartheta}_{k,l}$ with  $\tilde{\vartheta}_{k,l} \in [-\frac{\pi}{2}, \frac{\pi}{2}] $. In particular, for a uniform linear array (ULA)\footnote{Although ULA is considered in this work, the proposed approach can straightforwardly be extended for other array geometries, e.g., uniform rectangular array (URA), array-of-subarray (AoSA)~\cite{ummimoComMagYeLi,ummimoTareqOverview} or group-of-subarrays (GoSA)~\cite{elbir2021JointRadarComm}.}, the steering vector ${\mathbf{a}}(\vartheta_{k,l}) \in\mathbb{C}^{N_\mathrm{T}}  $ is defined as 
	\begin{align}
	\label{steeringVecPHY}
	\mathbf{a}(\vartheta_{k,l}) \hspace{-3pt}= \hspace{-3pt} \frac{1}{\sqrt{N_\mathrm{T}}} [1, e^{j 2\pi \frac{\bar{d}}{\lambda_c}  \vartheta_{k,l} },\cdots, e^{j 2\pi  \frac{\bar{d}}{\lambda_c}(N_\mathrm{T}-1) \vartheta_{k,l} }]^\textsf{T},
	\end{align}
	where $\lambda_c = \frac{c_0}{f_c}$, for which $c_0$ is speed of light and $f_c $ denotes the carrier frequency. Furthermore,  $\bar{d} = \frac{\lambda_c}{2}$ the half-wavelength array element spacing. {\color{black}Performing $M$-point DFT of the delay-$d$ channel in (\ref{channelTimeDomain}) yields 
		\begin{align}
		\mathbf{h}_k [m] &= \sum_{d=0}^{D-1}\bar{\mathbf{h}}_k(d) e^{-j \frac{2\pi m}{M}d}, \nonumber \\
		&= \sqrt{\frac{N_\mathrm{T}}{L}}\sum_{l =1}^{L} \left( \alpha_{k}^{l} \mathbf{a}(\vartheta_{k,l})  \sum_{d=0}^{D-1}  p(dT_s - \tau_{k,l})  e^{-j\frac{2 \pi m}{M} d } \right) , \nonumber \\
		& = \sqrt{\frac{N_\mathrm{T}}{L}}\sum_{l =1}^{L}  \alpha_{k}^{l} \mathbf{a}(\vartheta_{k,l})   e^{-j2 \pi \tau_{k,l} f_m } 
		\end{align}
		where $D\leq M$ is the CP length~\cite{channelEstThz,alkhateeb2016frequencySelective}. Define the $m$th subcarrier frequency as $f_m = f_c + \frac{B}{M}(m - 1 - \frac{M-1}{2}) $, where $B$ is the bandwidth. 
		
		Due to large bandwidth in THz systems, the channel direction $\vartheta_{k,l}$ becomes SD in spatial domain since $f_m \neq f_c$~\cite{thz_beamSplit,thz_channelEst_beamsplitPatternDetection_L_Dai,elbir2023Feb_BSA_OMP_WCL}. Therefore, the physical direction $\vartheta_{k,1}$ deviates to the spatial direction $\theta_{k,m,l}$, which is defined as 
		\begin{align}
		\label{physical_spatial_directions}
		\theta_{k,m,l} =  \frac{2 f_m}{c_0} \bar{d} \vartheta_{k,l} =  \frac{f_m}{f_c} \vartheta_{k,l} = \eta_m \vartheta_{k,l}.
		\end{align}
		Notice that $\vartheta_{k,l} = \theta_{k,m,l}$ if $f_m = f_c$, i.e., the narrowband scenario.	Therefore, we define the channel vector of the $k$th user at the $m$th subcarrier in frequency domain as
		\begin{align}
		\label{channelModel}
		\mathbf{h}_k[m]  =  
		\sqrt{\frac{N_\mathrm{T}}{L}}\sum_{l =1}^{L}  \alpha_{k}^{l} \mathbf{a}'(\theta_{k,m,l})   e^{-j2\pi\tau_{k,l} f_m },
		\end{align}
		where $\mathbf{a}'(\theta_{k,m,l}) \in \mathbb{C}^{N_\mathrm{T}}$ denotes the SD array steering vector under the effect of beam-split and it is  defined  as
		\begin{align}
		\label{steeringVec}
		\mathbf{a}'(\theta_{k,m,l}) \hspace{-3pt}&= \hspace{-3pt} \frac{1}{\sqrt{N_\mathrm{T}}} [1, e^{j 2\pi \frac{\bar{d}}{\lambda_m}  \vartheta_{k,l} },\cdots, e^{j 2\pi  \frac{\bar{d}}{\lambda_m}(N_\mathrm{T}-1) \vartheta_{k,l} }]^\textsf{T}, \nonumber \\
		& = \hspace{-3pt} \frac{1}{\sqrt{N_\mathrm{T}}} [1, e^{j \pi  \theta_{k,m,l} },\cdots, e^{j \pi (N_\mathrm{T}-1) \theta_{k,m,l} }]^\textsf{T}.
		\end{align}
	}
	Then, the channel model in (\ref{channelModel}) can  be expressed in a compact form as 
	\begin{align}
	\label{channelCompact}
	\mathbf{h}_k[m] = {\mathbf{A}}_k'[m]\tilde{\mathbf{x}}_k[m],
	\end{align}
	where ${\mathbf{A}}_k'[m] = [\mathbf{a}'(\theta_{k,m,1}),\cdots,\mathbf{a}'(\theta_{k,m,L})]\in \mathbb{C}^{N_\mathrm{T}\times L}$ and we have
	\begin{align}
	\tilde{\mathbf{x}}_k[m] = [\alpha_{k}^{1}e^{-j2\pi \tau_{k,1}f_m},\cdots,\alpha_{k}^{L}e^{-j2\pi \tau_{k,L}f_m}]^\textsf{T}\in \mathbb{C}^{L}.
	\end{align}


	In the following lemma, we show that the beam-split causes deviations in the channel directions such that maximum array gain is achieved at the beam-split deviated directions. 
	
	\begin{lemma}
		\label{lemmaArrayGain}
		Let $\mathbf{a}'(\theta_m)$ and $\mathbf{a}(\vartheta)$ be the actual and nominal steering vectors for an arbitrary physical direction $\vartheta$ and subcarrier $m\in \mathcal{M}$ as defined in (\ref{steeringVec}) and (\ref{steeringVecPHY}), respectively. Then, $\mathbf{a}(\theta_m)$ achieves the maximum array gain, i.e., $A_G(\vartheta,\theta_m) = \frac{ |\mathbf{a}^\textsf{H}(\vartheta)\mathbf{a}'(\theta_m)   |^2  }{N_\mathrm{T}^2} $, if $\theta_m = \eta_m \vartheta $.
	\end{lemma}

	\begin{IEEEproof} Please see Appendix~\ref{appendixAG}.
		
	\end{IEEEproof}


	Notice that both spatial and physical directions converges as $\theta_{k,m,l} \approx \vartheta_{k,l}$ if $f_m \approx f_c$, as shown in Fig.~\ref{fig_Diag}(c). This  allows one to design SI analog beamformers ($\mathbf{F}_\mathrm{RF}$) for $m\in\mathcal{M}$ in conventional wideband mmWave systems~\cite{beamSquint_FeiFei_Wang2019Oct,beamSquintGaoMagazine,beamSquintWang2019Nov}. However, beam-split implies that the physical directions $\vartheta_{k,l}$ deviate and completely split from the spatial directions $\theta_{k,m,l}$ as the system bandwidth widens, as shown in Fig.~\ref{fig_Diag}(b). Hence, we define the beam-split as the difference between the spatial and physical directions as
	\begin{align}
	\label{beamSplit}
	\Delta_{k,l}[m] = \theta_{k,m,l} - \vartheta_{k,l} = (\frac{f_m}{f_c} -1)\vartheta_{k,l},
	\end{align}
	which only depends on the frequency ratio $\frac{f_m}{f_c}$ and physical direction $\vartheta_{k,l}$.



	\subsection{Problem Formulation}
	\textcolor{black}{Due to the impact of beam-split, the spatial channel directions differ at each subcarrier. Therefore, a subcarrier-wise channel estimation is required to accurately estimate the channel.   }
	In downlink, the channel estimation stage is performed simultaneously by all the users  during channel training. Since the BS employs hybrid beamforming architecture, it activates only a single RF chain in each channel use to transmit the known pilot signals during channel acquisition~\cite{thz_beamSplit,thz_channelEst_beamsplitPatternDetection_L_Dai,channelEstThz}. Then, the BS employs $P$ beamformer vectors as $\tilde{\mathbf{F}}[m]= [\tilde{\mathbf{f}}_1[m],\cdots, \tilde{\mathbf{f}}_P[m]]\in \mathbb{C}^{N_\mathrm{T}\times P}$ ($|\tilde{\mathbf{f}}_p[m]| = 1/\sqrt{N}$) to send $P$ orthogonal pilots, $\tilde{\mathbf{S}}[m] = \mathrm{diag}\{\tilde{s}_1[m],\cdots, \tilde{s}_P[m]\}\in \mathbb{C}^{P\times P}$.  Hence, the pilot signals of $P$ frames are used to reconstruct the channel. By combining the received signals at the single-antenna users for $p =1,\cdots, P$, the $ P\times 1$ received signal at the $k$th user is 
	\begin{align}
	\label{receivedPilots}
	\bar{\mathbf{y}}_k[m] = \tilde{\mathbf{S}}[m]\bar{\mathbf{F}}[m]\mathbf{h}_k[m] + \mathbf{w}_k[m],
	\end{align}
	where we have  $\bar{\mathbf{F}} [m]= \tilde{\mathbf{F}}^\textsf{T}[m]\in \mathbb{C}^{P\times N_\mathrm{T}}$, $\bar{\mathbf{y}}_k[m] = [y_{k,1}[m],\cdots,y_{k,P}[m]]^\textsf{T}$ and ${\mathbf{w}}_k = [w_{k,1}[m],\cdots,w_{k,P}[m]]^\textsf{T}$. Defining the identical pilot signals for all subcarriers, i.e.,  $\mathbf{B} = \bar{\mathbf{F}}[m]$ and $\tilde{\mathbf{S}}[m] = \mathbf{I}_P$~\cite{angleDomain_CE_Fan2018Oct,alkhateeb2016frequencySelective,dovelos_THz_CE_channelEstThz2}, (\ref{receivedPilots}) becomes
	\begin{align}
	\label{receivedPilots2}
	\bar{\mathbf{y}}_k[m] = \mathbf{B}{\mathbf{h}}_k[m] + \mathbf{w}_k[m].
	\end{align}
	
	Estimating the channel from (\ref{receivedPilots2}) is performed via the traditional least squares (LS) and MMSE techniques as
	\begin{align}
	\label{ls_MMSE}
	\mathbf{h}_k^{\mathrm{LS}}[m] &= (\mathbf{B}^\textsf{H}\mathbf{B})^{-1}\mathbf{B}^\textsf{H}\bar{\mathbf{y}}_k[m]  , \nonumber \\
	\mathbf{h}_k^{\mathrm{MMSE}}[m] &= \big(\mathbf{R}_k^{-1}[m] + \mathbf{B}^\textsf{H} {\mathbf{R}}_k^{-1}[m]\mathbf{B}  \big)^{-1}\mathbf{B}^\textsf{H}\bar{\mathbf{y}}_k[m],
	\end{align}
	where $\mathbf{R}_k[m] = \mathbb{E}\{\mathbf{h}_k [m]\mathbf{h}_k^{\textsf{H}}[m]\}$ is the channel covariance matrix.  The estimators in (\ref{ls_MMSE}) may need additional priori information, e.g., covariance matrix or require $P\geq N_\mathrm{T}$, which can be computationally prohibitive for the computation of matrix inversions as well as entailing high  channel training overhead and time for large arrays. Further to that, the LS approach does not take into account the effect of beam-split, and prior information on the channel covariance is needed in MMSE estimator.
	
	Hence, our goal is to efficiently estimate $\mathbf{h}_k[m]$ for $m\in \mathcal{M}$ in the presence of beam-split when the number of received pilot signals is small. To this end, we exploit the sparsity of the THz channel and introduce an SBL-based approach in the following.
	
	%
	%
	
	\section{SBL for THz Channel Estimation}
	\label{sec:SBCE}
	The SBL method has been shown to be effective  for sparse signal reconstruction.	In this section, we first present the sparse THz signal model and introduce the proposed array perturbation model and an efficient approach for joint THz channel and beam-split estimation.
	
	\subsection{Sparse THz Channel Model}
	Due to employing UM number of antennas to compensate the significant path loss in THz frequencies, the channel is \textit{extremely-sparse}  ($L \ll N_\mathrm{T}$) in the beamspace~\cite{ummimoHBThzSVModel,ummimoComMagYeLi}. In order to exploit the sparsity of THz-band, the channel in (\ref{channelModel}) is expressed as
	\begin{align}
	\label{channelSparse}
	\mathbf{h}_k[m] = {\mathbf{D}}{\mathbf{x}}_k[m],
	\end{align}
	where ${\mathbf{x}}_k[m]\in \mathbb{C}^{N}$ is an $L$-sparse vector, i.e., $||{\mathbf{x}}[m]||_0=L$ and  ${\mathbf{D}} = [\mathbf{a}(\phi_1),\cdots,\mathbf{a}(\phi_N)]\in \mathbb{C}^{N_\mathrm{T}\times N}$ denotes an overcomplete dictionary  matrix composed of the steering vectors denoted by  $\mathbf{a}(\phi_n)\in \mathbb{C}^{N_\mathrm{T}}$ with $\phi_n = \frac{2n-N-1}{N}$ for $n = 1,\cdots, N$. Hence, the resolution of $\mathbf{D}$ is $\rho = 1/N$.

	%



	Next, we assume that the users collects only $P=N_\mathrm{RF}$ pilot signals, where $L\leq P\ll N_\mathrm{T}$. {\color{black}Further, we define $\mathbf{B}\in \mathbb{C}^{P\times N_\mathrm{T}}$ with $|[\mathbf{B}]_{i,j}| = \frac{1}{\sqrt{N_\mathrm{T}}}$ to represent  the SI beamformer matrix at the BS. Since there is a single RF chain at the user, the received pilot data during $P$ frames are stacked in a $P\times1$ vector as  
		\begin{align}
		\label{underdeterminedSystem}
		\mathbf{y}_k[m] =\mathbf{BF}{\mathbf{x}}_k[m] + \mathbf{w}_k[m].
		\end{align}
	}
	By exploiting the sparsity of ${\mathbf{x}}_k[m]$, the following $\ell_1$-norm sparse recovery (SR) problem is formulated, i.e., 
	\begin{align}
	\label{sparseRecovery}
	\hat{{\mathbf{x}}}_k[m] = &\arg  \minimize_{{\mathbf{x}}_k[m]} \hspace{5pt} || {\mathbf{x}}_k[m]  ||_1 \nonumber \\
	&	\subjectto \hspace{3pt}  ||  \mathbf{y}_k[m] - \mathbf{BF}{\mathbf{x}}_k[m] ||_2^2 \leq \epsilon,
	\end{align}
	where  the residual term $\epsilon$ is bounded by $\epsilon \leq\mu \sqrt{P + \kappa \sqrt{2P} }$, for which $\kappa$ corresponds to an adjustable tuning parameter in order to control the corruption due to noise~\cite{compressedSensing_SPM_Candes2008Mar,elbir2016Jul_Sparse_MutualCoupling}. One can solve (\ref{sparseRecovery}) for the $L$-sparse vector  $\hat{{\mathbf{x}}}_k[m]$, and the channel can be reconstructed as $\hat{\mathbf{h}}_k^{\text{SR}}[m] = {\mathbf{D}}\hat{{\mathbf{x}}}_k[m]$, which does not include beam-split correction.

	\subsection{Array Perturbation Based Beam-Split Model}
	Due to beam-split, the true physical channel directions are observed with deviations in the spatial domain at different subcarrier frequencies since frequency of the digital beamformer changes while the frequency of the analog beamformer remains the same. Therefore, conventional channel estimation approaches fail and the effect of beam-split should be taken into account. 
	To accurately mitigate the effect of beam-split, we approach the problem from an array calibration point of view, wherein the beam-split is modeled as an array perturbation, e.g., mutual coupling, gain/phase mismatch~\cite{elbir2016Jul_Sparse_MutualCoupling,sbl_doa_Liu2012Sep,sbl_CE_mmWave_Tang2018Jun}. Define $\mathbf{U}_k[m]\in \mathbb{C}^{N_\mathrm{T}\times N_\mathrm{T}} $  as the array perturbation matrix which maps the nominal array steering matrix, i.e., $\mathbf{A}_k$ to $\mathbf{A}_k'[m]$ which is perturbed due to SD beam-split.	Then, the channel model in (\ref{channelModel})  is 
	\begin{align}
	\label{arrayPerturbationChannel1}
	\mathbf{h}_k[m] &= \sqrt{\frac{N_\mathrm{T}}{L}}\sum_{l=1}^{L} \alpha_{k}^{l} \mathbf{U}_k[m] \mathbf{a}(\vartheta_{k,l})e^{-j2\pi \tau_{k,l}f_m},\nonumber\\
	&=  \mathbf{U}_k[m] {\mathbf{A}}_k \tilde{\mathbf{x}}_k[m],
	\end{align}
	where ${\mathbf{A}}_k = [{\mathbf{a}}(\vartheta_{k,1}),\cdots,{\mathbf{a}}(\vartheta_{k,L})]\in \mathbb{C}^{N_\mathrm{T}\times L}$. Thus, $\mathbf{U}_k[m]$ provides a linear transformation between the steering matrices corresponding to physical and spatial directions as 
	\begin{align}
	\label{transformationA_Atilde}
	{\mathbf{A}}_k'[m] = \mathbf{U}_k[m]{\mathbf{A}}_k.
	\end{align}
	Now, our aim is to rewrite the received signal $\mathbf{y}_k[m]$ as the linear combination of receiver output corresponding to the physical channel directions and some perturbation term.	{\color{black}Therefore, we first write $\mathbf{U}_k[m]{\mathbf{A}}_k\tilde{\mathbf{x}}_k[m]$ as 
		\begin{align}
		\label{ax_axQ}
		{\mathbf{A}}_k'[m]\tilde{\mathbf{x}}_k[m] = {\mathbf{A}}_k\tilde{\mathbf{x}}_k[m] + \bar{\mathbf{Q}}_k[m] \mathbf{u}_k[m],
		\end{align}
		where  $\bar{\mathbf{Q}}_k[m]$ is an $N_\mathrm{T}\times N_\mathrm{T}^2$ matrix with the following structure, i.e.,
		\begin{align}
		\label{QuAax}
		\bar{\mathbf{Q}}_k[m]\mathbf{u}_k[m] = \big({\mathbf{A}}_k'[m] - {\mathbf{A}}_k \big)\tilde{\mathbf{x}}_k[m],
		\end{align}
		where $\mathbf{u}_k[m] = \mathrm{vec}\{\mathbf{U}_k[m]\}\in \mathbb{C}^{N_\mathrm{T}^2}$, which includes the array perturbation terms in $\mathbf{U}_k[m]$. Now, we can rewrite the received signal model in (\ref{underdeterminedSystem}) by using (\ref{ax_axQ}) and (\ref{QuAax}) as 
		\begin{align}
		\mathbf{y}_k[m] &=\mathbf{B}{\mathbf{A}}_k'[m]\tilde{\mathbf{x}}_k[m] + \mathbf{w}_k[m] \nonumber \\
		&= \mathbf{B} \mathbf{U}_k[m]{\mathbf{A}}_k\tilde{\mathbf{x}}_k[m]    + \mathbf{w}_k[m]\nonumber\\
		\label{y_perturbed}& = \mathbf{B} ({\mathbf{A}}_k\tilde{\mathbf{x}}_k[m] + \bar{\mathbf{Q}}_k[m] \mathbf{u}_k[m]   ) + \mathbf{w}_k[m].
		\end{align}

		Now, our aim is to rewrite the received signal model with the combination of the nominal steering vectors (i.e., $\mathbf{B}\mathbf{A}_k $) and the array perturbation parameters (i.e., $\mathbf{u}$). Define $\tilde{\mathbf{P}}_k=  \mathbf{B}{\mathbf{A}}_k\in\mathbb{C}^{P\times L}$ and $\tilde{\mathbf{Q}}_k[m] = \mathbf{B}\bar{\mathbf{Q}}_k[m]\in\mathbb{C}^{P\times N_\mathrm{T}^2}$, then (\ref{y_perturbed}) becomes
		\begin{align}
		\label{y_perturbed_2}
		\mathbf{y}_k[m] = \tilde{\mathbf{P}}_k\tilde{\mathbf{x}}_k[m] + \tilde{\mathbf{Q}}_k[m] \mathbf{u}_k[m]   + \mathbf{w}_k[m].
		\end{align}
		
	}
	
	The perturbed array formulation in (\ref{y_perturbed_2}) explicitly shows the relationship between the received signal $\mathbf{y}_k[m]$ and the perturbation parameters corresponding to the beam-split $\mathbf{u}_k[m]$. Hence, one can minimize the fitting error between $\mathbf{y}_k[m]$ and $\tilde{\mathbf{P}}_k\tilde{\mathbf{x}}_k[m]$ to determine the physical channel directions as well as estimating the perturbation due to beam-split, as will be introduced in the following.
	
	\subsection{SBL}
	In the proposed SBL framework, we rewrite (\ref{y_perturbed_2}) in  an overcomplete form. Hence, only in this part, we first drop the subscript $(\cdot)_k$ and $[m]$ for notational simplicity, and discuss the channel estimation stage for multi-user multi-subcarrier case later. Next, we use $\mathbf{Q}\in \mathbb{C}^{P \times N_\mathrm{T}^2}$  instead of $\tilde{\mathbf{Q}}_k[m]$, and  define $\mathbf{P} = \mathbf{BF}\in \mathbb{C}^{P \times N}$ as the overcomplete version of $ \tilde{\mathbf{P}}_k\in \mathbb{C}^{P \times L}$  covering the whole angular domain with $\phi_n$, $n = 1,\cdots, N$. Also, we have $\mathbf{P}' =\mathbf{BUF}\in \mathbb{C}^{P\times N}$, i.e., the overcomplete version of the perturbed steering matrix $\mathbf{B}{\mathbf{A}}_k'[m] = \mathbf{BU}_k[m]{\mathbf{A}}_k\in \mathbb{C}^{P\times L}$. Then, (\ref{y_perturbed_2}) is expressed  as follows
	\begin{align}
	\label{y_perturbed_3}
	\mathbf{y} = {\mathbf{P}}{\mathbf{x}}+ {\mathbf{Q}} \mathbf{u}   + \mathbf{w},
	\end{align}
	where ${\mathbf{x}}$, $\mathbf{u}$ and $\mathbf{w}$ are $N\times 1$, $N_\mathrm{T}^2\times 1$ and $P\times 1$ vectors, respectively, as defined earlier with removed subscripts.
	
	Now, we introduce a new set of variables $\boldsymbol{\sigma} = [\sigma_1,\cdots,\sigma_N]^\textsf{T}$ as the variance of sparse vector $\mathbf{x}\sim \mathcal{CN}(\mathbf{0}, \boldsymbol{\Sigma})$, where $\boldsymbol{\Sigma} = \mathrm{diag}\{\boldsymbol{\sigma}\}$. Then, we derive the statistical dependence of the received signal $\mathbf{y}$ on the unknown parameters, i.e., the sparse vector $\mathbf{x}$, perturbation parameter $\mathbf{u}$, and the noise variance $\mu^2$. Toward this end, we look for the maximum likelihood estimate of these parameters to reconstruct  the channel vector from the support of $\mathbf{x}$ as $\hat{\mathbf{h}} = \mathbf{D}\hat{\mathbf{x}}$. 
	
	The likelihood function of the received signal in (\ref{y_perturbed_3}) is
	\begin{align}
	\label{pdf_y0}
	p(\mathbf{y}| \mathbf{x}; \mathbf{u},\mu^2 )&  = \frac{1}{\pi \mu^2N_\mathrm{T}} \exp\left\{\hspace{-1pt}-\frac{1}{\mu^2}\|\mathbf{y} \hspace{-3pt}-\hspace{-1pt} \mathbf{P}' \mathbf{x} \|_2^2   \right\} \nonumber \\
	& =\hspace{-3pt} \frac{1}{\hspace{-3pt} \pi \mu^2N_\mathrm{T}} \exp\left\{-\frac{1}{\mu^2}\|\mathbf{y} - \mathbf{Px} - \mathbf{Qu} \|_2^2   \hspace{-3pt}\right\}.
	\end{align}
	
	By using (\ref{pdf_y0}), we can write the probability density function (pdf) of $\mathbf{y}$ with respect to the hyperparameters $\boldsymbol{\sigma}$, $\mathbf{u}$ and $\mu^2$ as 
	\begin{align}
	\label{pdf_y}
	p(\mathbf{y}; \boldsymbol{\sigma}, \mathbf{u},\mu^2) &= \int  p (\mathbf{y}| \mathbf{x}; \mathbf{u}, \mu^2) \times p(\mathbf{x};\boldsymbol{\sigma}) d_{\mathbf{x}}\nonumber \\
	&= \frac{1}{|\pi \boldsymbol{\Pi}_\mathbf{y}|} \exp \left\{- \mathrm{Tr}\{\boldsymbol{\Pi}_\mathbf{y}^{-1}\mathbf{R}_\mathbf{y}  \}  \right\},
	\end{align} 
	where $\boldsymbol{\Pi}_\mathbf{y} \in \mathbb{C}^{P\times P}$ is the covariance of $\mathbf{y}$ as
	\begin{align}
	\label{cov_y}
	\boldsymbol{\Pi}_\mathbf{y}= \mathbb{E}\{\mathbf{y}\mathbf{y}^\textsf{H}\}  = \mathbf{P}'\boldsymbol{\Sigma} {\mathbf{P}'}^\textsf{H} + \mu^2\mathbf{I}_{ P},
	\end{align}
	and $\mathbf{R}_\mathbf{y} = \mathbf{y}\mathbf{y}^\textsf{H} \in \mathbb{C}^{P\times P}$.
	
	Define the unknown parameters in a hyperparameter set $\mathcal{S} = \{\boldsymbol{\sigma},\mathbf{u},\mu^2\}$. Then, $\mathcal{S}$ is estimated by maximizing the pdf in (\ref{pdf_y}), which is non-concave and computationally intractable due to the nonlinearities. Hence, we resort to the expectation-maximization (EM) algorithm, which iteratively converges to a local optimum~\cite{sbl1_tipping2001sparse,sbl2_Dempster1977Sep,sbl3_Radich1997Apr,sbl_doa_Liu2012Sep,sbl_CE_mmWave_Tang2018Jun}. Each EM iteration comprises two steps: E-step for \textit{inference}, and M-step for \textit{hyperparameter estimation}  by maximizing the Bayesian expectation of the complete probability $p(\mathbf{y}, \mathbf{x}, \boldsymbol{\sigma}, \mathbf{u},\mu^2 ) $. {\color{black}Specifically, the E-step involves the evaluation of the log-likelihood of the complete probability $p(\mathbf{y}, \mathbf{x}, \boldsymbol{\sigma}, \mathbf{u},\mu^2 ) $, i.e., $\mathbb{L}\left\{ \mathbf{y}, \mathbf{x}; \boldsymbol{\sigma}, \mathbf{u},\mu^2   \right\}$. Then in the M-step, it is maximized to estimate the hyperparameters $\{\boldsymbol{\sigma},\mathbf{u},\mu^2\}$. This is done by first computing the partial derivations of $\mathbb{L}\left\{ \mathbf{y}, \mathbf{x}; \boldsymbol{\sigma}, \mathbf{u},\mu^2  \right\}$ with respect to each of the hyperparameters and equating them to zero. Finally, we obtain the closed-form expressions to update each of the hyperparameters as discussed below.      }
	
	\subsubsection{E-Step (Inference)}
	{\color{black}In the E-step of the SBL algorithm, the log-likelihood of the probability $p(\mathbf{y}, \mathbf{x}, \boldsymbol{\sigma}, \mathbf{u},\mu^2 ) $ is evaluated, i.e, 
		\begin{align}
		\label{e_step1}
		\mathbb{L}\left\{\mathbf{y}, \mathbf{x}; \boldsymbol{\sigma}, \mathbf{u},\mu^2  \right\} &= \mathbb{E} \{ \ln p(\mathbf{y}, \mathbf{x}; \boldsymbol{\sigma}, \mathbf{u},\mu^2 ) \},
		\end{align}
		which can be written as
		\begin{align}
		\label{e_step1_2}
		\mathbb{L}\left\{ \mathbf{y}, \mathbf{x}; \boldsymbol{\sigma}, \mathbf{u},\mu^2   \right\}= \mathbb{E}\{ \ln p(\mathbf{y}| \mathbf{x}; \mathbf{u},\mu^2 )  +  \ln p( \mathbf{x}; \boldsymbol{\sigma}) \},
		\end{align}
		since $p(\mathbf{y}| \mathbf{x}; \mathbf{u},\mu^2 ) $ and $p( \mathbf{x}; \boldsymbol{\sigma}) $ do not depend on $\boldsymbol{\sigma}$ and $\{\mathbf{u}, \mu^2\}$, respectively~\cite{sbl_doa_Liu2012Sep,sbl_CE_mmWave_Tang2018Jun}. Then, the first term of (\ref{e_step1_2})  can be obtained from (\ref{pdf_y0}) as
		\begin{align}
		\label{e_step_3}
		\ln p(\mathbf{y}| \mathbf{x}; & \mathbf{u},\mu^2 )  \nonumber \\
		&=   \ln \left(\frac{1}{\hspace{-3pt} \pi \mu^2N_\mathrm{T}} \exp\left\{-\frac{1}{\mu^2}\|\mathbf{y} - \mathbf{Px} - \mathbf{Qu} \|_2^2   \hspace{-3pt}\right\} \right)  \nonumber\\
		& = -N_\mathrm{T} \ln \mu^2  - \mu^{-2}\| \mathbf{y} -  \mathbf{Px} - \mathbf{Qu}\|_2^2,
		\end{align}
		and the second term in (\ref{e_step1_2}) is given by 
		\begin{align}
		\label{e_step_4}
		\ln p( \mathbf{x}; \boldsymbol{\sigma}) =  - \sum_{n=1}^{N} ( \ln \sigma_n +  \frac{|x_n|^2}{\sigma_n } ),
		\end{align}
		since $\mathbf{x}\sim \mathcal{CN}(\mathbf{0}, \boldsymbol{\Sigma})$~\cite{sbl_varX_Wu2015Dec}. Combining (\ref{e_step_3}) and (\ref{e_step_4}),  (\ref{e_step1}) becomes 
		\begin{align}
		\label{e_step2}
		&\mathbb{L}\{ \mathbf{y},  \mathbf{x}; \boldsymbol{\sigma}, \mathbf{u},\mu^2   \} \nonumber \\
		&= \mathbb{E}\{   -N_\mathrm{T} \ln \mu^{2} - \mu^{-2} \|\mathbf{y} - \mathbf{Px} - \mathbf{Qu} \|_2^2\nonumber \\
		&\hspace{20pt}  - \sum_{n=1}^{N} ( \ln \sigma_n +  \frac{|x_n|^2}{\sigma_n } )   \}.
		\end{align}
		
	}

	\subsubsection{M-Step (Hyperparameter Estimation)}
	\label{sec:Mstep}
	Now, we consider estimating the hyperparameter set $\mathcal{S}$ by maximizing $\mathbb{L}\{ \mathbf{y},  \mathbf{x}; \boldsymbol{\sigma}, \mathbf{u},\mu^2   \}$. To this end, we first compute the partial derivatives of (\ref{e_step2}) with respect to the unknown parameters as
	\begin{align}
	&\frac{\partial}{\partial \mathbf{u}} \mathbb{L}\{ \mathbf{y},  \mathbf{x}; \boldsymbol{\sigma}, \mathbf{u},\mu^2   \}\nonumber \\
	&= -2 \mu^{-2} \bigg( \mathbb{E}\{ \mathbf{Q}^\textsf{H}\mathbf{Q} \} \mathbf{u} - \mathbb{E}\{ \mathbf{Q}^\textsf{H} (\mathbf{y} - \mathbf{P} \mathbf{x})  \} \bigg), \\
	&\frac{\partial}{\partial \mu^2} \mathbb{L}\{ \mathbf{y},  \mathbf{x}; \boldsymbol{\sigma}, \mathbf{u},\mu^2  \}
	\hspace{-3pt}=\hspace{-3pt} - \frac{N_\mathrm{T}}{\mu^2}\hspace{-3pt} + \hspace{-3pt} \frac{1}{\mu^4} \mathbb{E}\{ \|\mathbf{y} - \mathbf{P}' \mathbf{x} \|_2^2  \}, \\
	&\frac{\partial}{\partial {\sigma}_n} \mathbb{L}\{ \mathbf{y},  \mathbf{x}; \boldsymbol{\sigma}, \mathbf{u},\mu^2   \}
	= - \frac{1}{\sigma_n} + \frac{1}{\sigma_n^2} \mathbb{E}\{ |{x_n}|^2  \},
	\end{align}
	where $n = 1,\cdots,N$. 
	By setting the above derivatives to zero, we can get the following expressions, i.e.,
	\begin{align}
	\label{partial_mu}\mu^2 &= \frac{1}{N_\mathrm{T}}  \mathbb{E}\{  \|\mathbf{y} - \mathbf{P}' \mathbf{x} \|_2^2  \}, \\
	\sigma_{n}& =   \mathbb{E}\{ |x_n|^2\}  , \label{partial_sigma} \\
	\label{partial_u}	\mathbf{u} &= {\color{black}\mathbb{E}\big\{ \mathbf{Q}^\textsf{H}\mathbf{Q}    \}^{-1} }\mathbb{E}\big\{ \mathbf{Q}^\textsf{H} (\mathbf{y} - \mathbf{Px})\big\}. 
	\end{align}

	{\color{black}The expression  $ \mathbb{E}\{  \|\mathbf{y} - \mathbf{P}'\mathbf{x}  \|_2^2  \}$ in (\ref{partial_mu}) can be written as 
		\begin{align}
		\label{y_Px_posterior1}
		\mathbb{E}\{  \|\mathbf{y} - \mathbf{P}' \mathbf{x} \|_2^2  \} =  ||\mathbf{y} - {\mathbf{P}'} \mathbf{z}  ||_2^2 + \mathrm{Tr}\{\mathbf{P}' \boldsymbol{\Pi} {\mathbf{P}'}^\textsf{H} \},
		\end{align}
		by utilizing  the posterior density of $\mathbf{x}$ is  $p(\mathbf{x}| \mathbf{y}) \sim \mathcal{CN}(\mathbf{z}, \boldsymbol{\Pi})$, where  $\mathbf{z} = \mu^{-2}\boldsymbol{\Pi} {\mathbf{P}'}^{\textsf{H}}\mathbf{y} \in \mathbb{C}^{N}$ and $\boldsymbol{\Pi} = \boldsymbol{\Sigma} -  \boldsymbol{\Sigma}{\mathbf{P}'}^\textsf{H} \boldsymbol{\Pi}_\mathbf{y}^{-1}\mathbf{P}' \boldsymbol{\Sigma}    \in\mathbb{C}^{N\times N} $. Please see Appendix~\ref{appendix_Posterior1} for the proof.

	}

	Furthermore, the $(i,j)$th element of the first expectation term  in (\ref{partial_u}) is computed as
	\begin{align}
	[\mathbb{E}\{ \mathbf{Q}^\textsf{H}\mathbf{Q}    \}]_{ij} &=  \mathbb{E}\{ \mathbf{x}^\textsf{H} \mathbf{D}^\textsf{H} \mathbf{M}_i^\textsf{H}\mathbf{M}_j \mathbf{D}\mathbf{x}  \}  \nonumber \\
	&= \mathrm{Tr}\{\mathbf{M}_i^\textsf{H}\mathbf{M}_j \mathbf{D}  \mathbb{E}\{ \mathbf{x}\mathbf{x}^\textsf{H}\} \mathbf{D}^\textsf{H}  \},	\label{QQ}
	\end{align}
	where $\mathbf{M}_i\in \mathbb{C}^{P\times N_\mathrm{T}}$ denotes the derivative of $\mathbf{BU}$ with respect to $u_i$ as 
	\begin{align}
	\mathbf{M}_i = \frac{\mathbf{B}\partial\mathbf{U}}{\partial u_i},
	\end{align}  for $i = 1,\cdots,N_\mathrm{T}^2$~\cite{sbl_doa2_Liu2013May}. \textcolor{black}{In other words,  the $i$th column of $\mathbf{Q}$ is given by 
		\begin{align}
		[\mathbf{Q}]_i = \mathbf{M}_i \mathbf{D}\mathbf{x}.
		\end{align}} Further,
	using (\ref{y_perturbed_3}), we can compute $ \mathbb{E}\{ \mathbf{x}\mathbf{x}^\textsf{H}\}$ in (\ref{QQ})  as
	\begin{align}
	\mathbb{E}\{ \mathbf{x}\mathbf{x}^\textsf{H}\} = \boldsymbol{\Sigma}= \mathbf{z} \mathbf{z}^\textsf{H} + \boldsymbol{\Pi}  .
	\end{align} Then, (\ref{QQ}) becomes
	\begin{align}
	[\mathbb{E}\{ \mathbf{Q}^\textsf{H}\mathbf{Q}    \}]_{ij}=\mathrm{Tr}\{\mathbf{M}_i^\textsf{H}\mathbf{M}_j\mathbf{D} (\mathbf{z} \mathbf{z}^\textsf{H} + \boldsymbol{\Pi} ) \mathbf{D}^\textsf{H}  \} .
	\end{align}
	Also, the $i$th element of the expectation {\color{black} $\mathbb{E}\big\{   \mathbf{Q}^\textsf{H} (\mathbf{y} - \mathbf{Px})\big\}$ in (\ref{partial_u})} is obtained as
	\begin{align}
	\label{Qy_Px}
	[\mathbb{E}\{ \mathbf{Q}^\textsf{H} (\mathbf{y} - \mathbf{Px}) \}]_i &= [\mathbb{E}\{ \mathbf{Q}^\textsf{H} \mathbf{y} \}]_i -  [\mathbb{E}\{ \mathbf{Q}^\textsf{H}  \mathbf{Px} \}]_i \nonumber \\
	&= \mathrm{Tr}\{\mathbf{M}_i^\textsf{H}\mathbf{y}\mathbf{z}^\textsf{H} \mathbf{D}^\textsf{H}  \}  - \mathrm{Tr}\{\mathbf{M}_i^\textsf{H} \mathbf{P} (\mathbf{z}\mathbf{z}^\textsf{H} \nonumber\\
	&\hspace{20pt} + \boldsymbol{\Pi} \mathbf{P}^\textsf{H} )\mathbf{D}^\textsf{H} \},
	\end{align}
	for $i = 1,\cdots, N_\mathrm{T}^2$. 
	
	Finally, we write the expectation-free expressions for the EM algorithm in  closed-form as
	\begin{align}
	\label{partial_mu2}\mu^2 &= \frac{1}{N_\mathrm{T}}  ||\mathbf{y} - {\mathbf{P}'} \mathbf{z}  ||_2^2 + \mathrm{Tr}\{\mathbf{P}' \boldsymbol{\Pi} {\mathbf{P}'}^\textsf{H} \}, \\
	\sigma_{n}& =   | z_n|^2, \label{partial_sigma2} \\
	\label{partial_u2}	{u}_i &=   \mathrm{Tr}\{\mathbf{M}_i^\textsf{H}\mathbf{M}_j\mathbf{D} (\mathbf{z}\mathbf{z}^\textsf{H} + \boldsymbol{\Pi} ) \mathbf{D}^\textsf{H}  \} \bigg(\mathrm{Tr}\{\mathbf{M}_i^\textsf{H}\mathbf{y}\mathbf{z}^\textsf{H} \mathbf{D}^\textsf{H}  \}  \nonumber\\
	&\hspace{20pt}- \mathrm{Tr}\{\mathbf{M}_i^\textsf{H} \mathbf{P} (\mathbf{z}\mathbf{z}^\textsf{H} + \boldsymbol{\Pi} \mathbf{P}^\textsf{H} )\mathbf{D}^\textsf{H} \}\bigg),
	\end{align}
	for $n= 1,\cdots, N$ and  $i,j = 1,\cdots,N_\mathrm{T}^2$.

	%
	%
	
	By computing (\ref{partial_mu2})-(\ref{partial_u2}) iteratively, one can estimate the parameters $\boldsymbol{\sigma}$, $\mathbf{u}$ and $\mu^2$. Since the EM algorithm is proved to be convergent~\cite{sbl1_tipping2001sparse,sbl2_Dempster1977Sep,sbl_CE_mmWave_Tang2018Jun,sbl_doa_Liu2012Sep}, most of the terms in $\boldsymbol{\sigma}$ and $\mathbf{x}$ converge to $0$, yielding to a sparse profile.
	

	\subsection{Low-Complexity Approach for Array Perturbation Update}
	Updating $\mathbf{u}$ involves the computation of $(\mathbf{Q}^\textsf{H}\mathbf{Q})^{-1}\in \mathbb{C}^{N_\mathrm{T}^2\times N_\mathrm{T}^2}$ and $\mathbf{Q}(\mathbf{y} - \mathbf{Px})\in \mathbb{C}^{N_\mathrm{T}^2} $ in  (\ref{partial_u})  which are computationally prohibitive due to large number of unknown parameters, i.e., ${N}_\mathrm{T}^2$, especially when the number of antennas is high. Therefore, we propose a low-complexity approach in the following.
	
	\textcolor{black}{Instead of employing an $N_\mathrm{T}\times N_\mathrm{T}$ full matrix $\mathbf{U}_k[m]$ in (\ref{arrayPerturbationChannel1}) to represent the array perturbations, we exploit the LoS-dominant feature of the THz channel, we assume that $L = 1$ by neglecting the NLoS paths, which are approximately $10$ dB weaker than the LoS paths~\cite{ummimoTareqOverview,teraMIMO}. This allows us to model $\mathbf{U}_k[m]$ as a diagonal matrix $\mathbf{C}_k[m]\in \mathbb{C}^{N_\mathrm{T}\times N_\mathrm{T}}$ instead of a full matrix. Rewrite (\ref{transformationA_Atilde}) for $L=1$ as
		\begin{align}
		\label{transformationA_Atilde_2}
		{\mathbf{a}}_k'[m] = \mathbf{C}_k[m]{\mathbf{a}}_k,
		\end{align}
		where ${\mathbf{a}}_k'[m]\in \mathbb{C}^{N_\mathrm{T}}$ and ${\mathbf{a}}_k\in \mathbb{C}^{N_\mathrm{T}}$ correspond to the perturbed and nominal array steering vectors for the LoS paths, respectively. In (\ref{transformationA_Atilde_2}), $\mathbf{C}_k[m] = \mathrm{diag} (\mathbf{c}_k[m]) $ is an ${N_\mathrm{T}\times N_\mathrm{T}}$ 
		diagonal transformation matrix.} $\mathbf{c}_k[m]$ corresponds to the perturbation due to beam-split and its $i$th entry is defined as
	\begin{align}
	[\mathbf{c}_k[m]]_{i} = e^{j\pi (i-1) \Delta_k[m] },
	\end{align}
	for $i = 2,\cdots, N_\mathrm{T}$ and $[\mathbf{c}_k[m]]_1 = 1$. Notice that this approach involves only a single  unknown parameter, i.e., $\Delta_k[m]$ to perform the update from $\mathbf{a}_k[m]$ to $\mathbf{a}_k'[m]$ whereas $N_\mathrm{T}^2$ parameters are involved in (\ref{partial_u}). Furthermore, the transformation in (\ref{transformationA_Atilde_2}) also allows us to accurately estimate the beam-split as presented in the following lemma.
	


	\begin{lemma} 
		\label{lem:1}
		Let ${\theta}_{k,m}$ be the spatial channel direction, then the beam-split introduced at the $m$th subcarrier is uniquely recovered as
		\begin{align}
		\label{lemma1}
		\Delta_k[m] &= \frac{1}{N_\mathrm{T}-1}\sum_{i = 2}^{N_\mathrm{T}} \frac{\angle [\mathbf{c}_k[m]]_i }{\pi (i-1)},
		\end{align}
		where $\angle [\mathbf{c}_k[m]]_i$ is the  angle of $[\mathbf{c}_k[m]]_i$. 
	\end{lemma}
	\begin{IEEEproof} Consider the steering vector $\mathbf{a}({\theta}_{k,m})$. Then, using (\ref{beamSplit}), the $i$th entry of $\mathbf{a}({\theta}_{k,m})$ is given by
		\begin{align}
		\label{a_i}
		[\mathbf{a}({\theta}_{k,m})]_i = e^{j\pi (i-1){\theta}_{k,m}  }.
		\end{align}
		Now, we find the angle of $[\mathbf{a}({\theta}_{k,m})]_i$ as $\Omega_{k,i}[m] = \angle [\mathbf{a}({\theta}_{k,m})]_i$ for $i = 1,\cdots,N_\mathrm{T}$. Then, we compute the unwrapped angles as $\boldsymbol{\Omega}_k [m] = \mathrm{unwrap}\{ [\Omega_{k,1}[m],\cdots, \Omega_{k,N_\mathrm{T}}[m]]^\textsf{T} \}$ to ensure not losing information due to the $[-\pi,\pi]$ periodicity of the exponential\footnote{The unwrapped angle information is  easily obtained via the \texttt{unwrap} command in MATLAB.}. Using $\boldsymbol{\Omega}_k[m]$ and (\ref{beamSplit}), the $i$th element of the steering vector corresponding to the physical direction ${\vartheta}_k = \frac{f_c}{f_m} \theta_{k,m}$  becomes
		\begin{align}
		\label{aiprime}
		[\mathbf{a}({\vartheta}_k)]_i = e^{j \Omega_{k,i}[m] \frac{f_c}{f_m}  }.
		\end{align}
		Substituting (\ref{aiprime}) and  (\ref{a_i}) into (\ref{transformationA_Atilde_2}) yields $[\mathbf{a}({\theta}_{k,m}) ]_i= c_{k,i} [m] 	[\mathbf{a}({\vartheta_k})]_i,$ where the angle of $[\mathbf{c}_{k}[m]]_i$ is given by
		\begin{align}
		\angle [\mathbf{c}_k[m]]_i &=  \Omega_{k,i}[m] (1 - \frac{f_c}{f_m} ) \nonumber \\
		&= \pi (i-1) \theta_{k,m} (\frac{\theta_{k,m} - \vartheta_{k}}{\theta_{k,m}} ) \nonumber \\
		& = \pi (i-1) \Delta_k[m].
		\end{align}
		Then,  taking average of $\angle [\mathbf{c}_k[m]]_i$ for $i = 2,\cdots,N_\mathrm{T}$ and leaving alone $\Delta_k[m]$  leads to  (\ref{lemma1}).
		
	\end{IEEEproof}

	Lemma~\ref{lem:1} allows us to implement the SBL iterations with significantly lower complexity. We present the algorithmic steps of SBCE in Algorithm~\ref{alg:SBCE}. In particular, the SBCE is initialized with ($t = 0$) $\sigma^{(t)}= \mathbf{1}_N$, $\mathbf{C}_k^{(t)}[m] = \mathbf{I}_{{N}_\mathrm{T}}$ and ${\mu^2}^{(t)} =0$. Then, the covariances ${\boldsymbol{\Pi}_{\mathbf{y}}}^{(t)}$ and ${\boldsymbol{\Pi}_{\mathbf{x}}}^{(t)}$ are computed. As the SBCE iterates, $\mathbf{C}_k^{(t)}[m]$ involves the array perturbations due to beam-split and providing a linear transformation between the nominal and actual steering vectors as $\mathbf{a}(\frac{f_m}{f_c}\phi_{n^\star}^{(t)}) = \mathbf{C}^{(t)}\mathbf{a}(\phi_{n^\star}^{(t)})$. \textcolor{black}{Once the SBCE converges, the beam-split and physical direction $\hat{\vartheta}_{k}$ are estimated. Then, the channel support from (\ref{channelSparse}) is reconstructed as $\hat{{{x}}}_k[m] = (\mathbf{B}	\mathbf{a}(\hat{\vartheta}_{k}))^\dagger \mathbf{y}_k[m]$ for $L=1$, and the estimated channel is obtained as $\mathbf{\hat{\mathbf{h}}}_k[m] = \mathbf{a}(\hat{\vartheta}_{k}) \hat{{{x}}}_k[m]$.}
	In the following, we discuss refining the estimated physical directions.


	\begin{algorithm}[t]
		\begin{algorithmic}[1] 
			\caption{ \bf SBCE}
			\Statex {\textbf{Input:}  $\mathbf{D}$, $\mathbf{B}$,  $\epsilon_\mathrm{TH}$, $f_c$, $\mathbf{y}_k[m]$, $f_m$ for $m\in \mathcal{M}$.} \label{alg:SBCE}
			\Statex {\textbf{Output:} $\hat{\mathbf{h}}_k[m]$, $\hat{\vartheta}_{k}$, $\hat{\Delta}_{k}[m]$.}
			\State \textbf{for} $k \in\mathcal{K}$
			\State \textbf{for} $m\in \mathcal{M}$
			\State  Initialize: $t\hspace{-2pt} =\hspace{-2pt} 0$, $\boldsymbol{\sigma}^{(t)} \hspace{-2pt}=\hspace{-2pt} \mathbf{1}_{N}$, $\mathbf{C}_k^{(t)}[m] \hspace{-2pt}= \hspace{-2pt}\mathbf{I}_{N_\mathrm{T}}$ and ${\mu^2}^{(t)} \hspace{-2pt}=\hspace{-2pt} 0$.
			\State  $\mathbf{y} = \mathbf{y}_k[m]$,  $\mathbf{C}^{(t)} = \mathbf{C}_k^{(t)}[m]$, $\mathbf{P}'^{(t)}= \mathbf{B} \mathbf{C}^{(t)}\mathbf{D} $.
			\State  $\texttt{flag}= \mathrm{true}$.
			\State \textbf{while} $\texttt{flag}= \mathrm{true}$ 
			\State \indent $\boldsymbol{\Sigma}^{(t)} = \mathrm{diag}\{\boldsymbol{\sigma}^{(t)}\}$.
			\State \indent ${\boldsymbol{\Pi}_{\mathbf{y}}}^{(t)} =  \mathbf{P}'^{(t)}\boldsymbol{\Sigma}^{(t)} {\mathbf{P}'^{(t)}}^\textsf{H} + {\mu^2}^{(t)}\mathbf{I}_{P}$.
			\State \indent $\boldsymbol{\Pi}_{\mathbf{x}}^{(t)} = (\boldsymbol{\Sigma}^{(t)} -  \boldsymbol{\Sigma}^{(t)}{\mathbf{P}'^{(t)}}^\textsf{H} {\boldsymbol{\Pi}_{\mathbf{y}}^{(t)}}^{-1}\mathbf{P}'^{(t)} \boldsymbol{\Sigma}^{(t)}   )^{-1}$.
			\State \indent $t \leftarrow t + 1$.
			\State \indent $\mathbf{z}^{(t)} = {\mu^2}^{(t)}{\boldsymbol{\Pi}_{\mathbf{x}}}^{(t)} {\mathbf{P}'^{(t)}}^{\textsf{H}}\mathbf{y}$.
			\State \indent Update $\sigma_{{n}}^{(t)} =  |z_{n}^{(t)}|^2$.
			\State \indent Find $\phi_{n^\star}^{(t)}$ from $n^\star  = \argmax_n |z_n^{(t)}
			|^2$.
			\State \indent Construct $\mathbf{a}(\phi_{n^\star}^{(t)})$.
			\State \indent Update $\mathbf{C}^{(t)}$ as $\mathbf{a}(\frac{f_m}{f_c}\phi_{n^\star}^{(t)}) = \mathbf{C}^{(t)}\mathbf{a}(\phi_{n^\star}^{(t)})$.
			\State \indent Update $\mathbf{P}'^{(t)}= \mathbf{B} \mathbf{C}^{(t)}\mathbf{D} $.
			
			\State \indent Update ${\mu^2}^{(t)} = \frac{1}{N_\mathrm{T}} ||\mathbf{y} - {\mathbf{P}'^{(t)}}^\textsf{H} \mathbf{z}^{(t)}  ||_2^2 + $\par \indent $ \mathrm{Tr}\{\mathbf{P}'^{(t)}{\boldsymbol{\Pi}}^{(t)} {\mathbf{P}'^{(t)}}^\textsf{H} \}$.
			
			\State \indent \textbf{if} $|| \boldsymbol{\sigma}^{(t)} - \boldsymbol{\sigma}^{(t-1)}||_2/ ||\boldsymbol{\sigma}^{(t)} ||_2 < \epsilon_\mathrm{TH}$
			\State \indent \indent \texttt{flag} = $\mathrm{false}$.
			\State \indent \textbf{end if}
			\State \textbf{end while}
			\State $ \mathbf{y}_k[m] = \mathbf{y}$,  $\mathbf{C}_k[m] = \mathbf{C}^{(t)} $, $\bar{\vartheta}_{k} = \phi_{n^\star}^{(t)}$.
			\State Insert $\bar{\vartheta}_{k}$ into (\ref{cov_y_2})-(\ref{refined_direction_find}), find the refined directions $\hat{\vartheta}_{k}$.
			\State Beam-split estimate: $\hat{\Delta}_k[m] = \frac{1}{N_\mathrm{T}-1}\sum_{i = 2}^{N_\mathrm{T}}  \{\frac{\angle[\mathbf{c}_k[m]]_i}{\pi (i-1) }   \}  $.
			\State {\color{black}Support estimate $\hat{{{x}}}_k[m] = (\mathbf{B}	\mathbf{a}(\hat{\vartheta}_{k}))^\dagger \mathbf{y}_k[m]$.   }
			\State Channel estimate: $\mathbf{\hat{\mathbf{h}}}_k[m] = \mathbf{a}(\hat{\vartheta}_{k}) \hat{{{x}}}_k[m]$.
			\State \textbf{end for}
			\State \textbf{end for}
		\end{algorithmic} 
	\end{algorithm}

	\subsection{Refined Direction Estimation}
	\label{sec:Refine}
	The estimation accuracy of the direction estimates obtained via the support of $\mathbf{x}$ is subject to the angular resolution of the selected grid, i.e., $\rho = \frac{1}{N}$. Hence, once the EM algorithm terminates, we use the resulting estimates and search for the refined directions by employing a finer grid~\cite{sbl_doa_Liu2012Sep,sbl_CE_mmWave_Tang2018Jun}. Define the finer grid for the $l$th physical direction $\vartheta_l$ as ${\Phi}_l = [\vartheta_l - \varphi, \vartheta_l + \varphi ]$, where the grid interval is $2\varphi $. Then, we rewrite the covariance of the received signal $\boldsymbol{\Pi}_\mathbf{y}$ as
	\begin{align}
	\label{cov_y_2}
	\boldsymbol{\Pi}_\mathbf{y} = \boldsymbol{\Pi}_{\mathbf{y} \backslash l} + \eta_l \mathbf{g}'(\vartheta_l){\mathbf{g}'}^\textsf{H}(\vartheta_l),
	\end{align}
	where $\mathbf{g}'(\vartheta_l) = \mathbf{B}\mathbf{C}\mathbf{a}(\vartheta_l)\in \mathbb{C}^{P}$ and  $\eta_l$ is the power of the $l$th signal. $\boldsymbol{\Pi}_{\mathbf{y}\backslash l}$ denotes to the covariance matrix in (\ref{cov_y}) by excluding the columns of $\mathbf{P}'$ corresponding to $\vartheta_l$. By substituting (\ref{cov_y_2}) into (\ref{pdf_y}), we can obtain the refined directions by maximizing  (\ref{pdf_y0}), wherein we substitute $\eta_l$ and $\vartheta_l$ as $	\{\hat{\vartheta}_l, \hat{\eta}_l\}  = \arg \max_{\eta_l,\vartheta_l} f(\eta_l,\vartheta_l),$ 	where $f(\eta_l,\vartheta_l)$ is 
	\begin{align}
	\label{refined_problem2}
	&f(\eta_l,\vartheta_l) = \frac{1}{|\pi [\boldsymbol{\Pi}_{\mathbf{y}\backslash l} + \eta_l \mathbf{g}'(\vartheta_l){\mathbf{g}'}^\textsf{H}(\vartheta_l)]|}  \nonumber \\
	&\hspace{15pt}\times \exp \left\{- \mathrm{Tr}\left\{  [\boldsymbol{\Pi}_{\mathbf{y}\backslash l} + \eta_l \mathbf{g}'(\vartheta_l){\mathbf{g}'}^\textsf{H}(\vartheta_l)]^{-1}\mathbf{R}_\mathrm{y} \right\}  \right\},
	\end{align}
	for which the equivalent minimization problem is given by $\{\hat{\vartheta}_l, \hat{\eta}_l\} = \arg \min_{\eta_l,\vartheta_l} \bar{f}(\eta_l,\vartheta_l)$, where 
	\begin{align}
	\label{refined_problem3}
	\bar{f}(\eta_l,\vartheta_l) = &\ln |\pi [\boldsymbol{\Pi}_{\mathbf{y}\backslash l} + \eta_l \mathbf{g}'(\vartheta_l){\mathbf{g}'}^\textsf{H}(\vartheta_l)]|   \nonumber\\
	&\hspace{20pt}+   \mathrm{Tr}\left\{  [\boldsymbol{\Pi}_{\mathbf{y}\backslash l} + \eta_l \mathbf{g}'(\vartheta_l){\mathbf{g}'}^\textsf{H}(\vartheta_l)]^{-1}\mathbf{R}_\mathrm{y} \right\}  ,
	\end{align}
	
	To simplify (\ref{refined_problem3}), we first set the derivative $\frac{\partial \bar{f}(\eta_l,\vartheta_l)}{\partial \eta_l}$ to zero, which yields
	\begin{align}
	\label{eta}
	\hat{\eta}_l = \frac{ {\mathbf{g}'}^\textsf{H}(\vartheta_l) \boldsymbol{\Pi}_{\mathbf{y} \backslash l}^{-1} (\mathbf{R}_\mathbf{y} - \boldsymbol{\Pi}_{\mathbf{y} \backslash l}) \boldsymbol{\Pi}_{\mathbf{y} \backslash l}^{-1} \mathbf{g}'(\vartheta_l) }{ [{\mathbf{g}'}^\textsf{H}(\vartheta_l) \boldsymbol{\Pi}_{\mathbf{y} \backslash l}^{-1}  {\mathbf{g}'}(\vartheta_l)] },
	\end{align}
	and insert (\ref{eta}) into $\frac{\partial \bar{f}(\eta_l,\vartheta_l)}{\partial \vartheta_l} =0$. Then, we get
	\begin{align}
	\label{f_nice}
	&\mathrm{Re}\{ {\mathbf{g}'}^\textsf{H}(\vartheta_l)\boldsymbol{\Pi}_{\mathbf{y} \backslash l}^{-1} [ {\mathbf{g}'}(\vartheta_l){\mathbf{g}'}^\textsf{H}(\vartheta_l) \boldsymbol{\Pi}_{\mathbf{y} \backslash l}^{-1}\mathbf{R}_\mathbf{y} \nonumber \\ 
	&\hspace{20pt}- \mathbf{R}_\mathbf{y}\boldsymbol{\Pi}_{\mathbf{y} \backslash l}^{-1}{\mathbf{g}'}(\vartheta_l){\mathbf{g}'}^\textsf{H}(\vartheta_l)     ]  \boldsymbol{\Pi}_{\mathbf{y} \backslash l}^{-1}  {\dot{\mathbf{g}}'}(\vartheta_l) \} = 0,
	\end{align}
	where ${\dot{\mathbf{g}}'}(\vartheta_l) = \frac{\partial {\mathbf{g}'}(\vartheta_l)}{\partial \vartheta_l}$. Using (\ref{f_nice}), we can write the refined direction estimation problem as
	\begin{align}
	\label{refined_direction_find}
	&\hat{\vartheta}_l = \arg \max_{\vartheta_l} | \mathrm{Re}\{ {\mathbf{g}'}^\textsf{H}(\vartheta_l)\boldsymbol{\Pi}_{\mathbf{y} \backslash l}^{-1} [ {\mathbf{g}'}(\vartheta_l){\mathbf{g}'}^\textsf{H}(\vartheta_l) \boldsymbol{\Pi}_{\mathbf{y} \backslash l}^{-1}\mathbf{R}_\mathbf{y} \nonumber \\ 
	&\hspace{20pt}- \mathbf{R}_\mathbf{y}\boldsymbol{\Pi}_{\mathbf{y} \backslash l}^{-1}{\mathbf{g}'}(\vartheta_l){\mathbf{g}'}^\textsf{H}(\vartheta_l)     ]  \boldsymbol{\Pi}_{\mathbf{y} \backslash l}^{-1}  {\dot{\mathbf{g}}'}(\vartheta_l) \} |^{-1}.
	\end{align}

	\subsection{Computational Complexity}
	The complexity of SBCE is mainly due to the matrix computations in Algorithm~\ref{alg:SBCE}, which are $O(PN^2+NP^2)$ (Step 8),
	$O(2N^2P + NP^2 +2N^3 + P^3)$ (Step 9),
	$O(N^2P + NP)$ (Step 11),
	$O(P^2 N_\mathrm{T}^3 N)$ (Step 16) and
	$O(NP + PN^2 + P^2N)$ (Step 17), respectively. Hence, the overall complexity order is $O(T \big[ PN (5N^2 + 3P^2 +PN_\mathrm{T}^3 + 2) + 2P^3 + 3N^3\big])$. In addition, the complexity of the direction refinement process in Sec.~\ref{sec:Refine} is proportional to the number of grid points to compute (\ref{refined_direction_find}).

	\section{Near-field Considerations}
	\label{sec:nearField}
	Due to operating at high frequencies as well as employing extremely small array aperture, THz-band transmission may encounter near-field phenomenon for close-proximity users. Specifically, the far-field model involves the reception of the transmitted signal  at the users as plane-wave whereas the plane wavefront is spherical in the near-field if the transmission range is shorter than the Fraunhofer distance, i.e., $R = \frac{2G^2 f_c}{c_0}  $ where $G$ is the array aperture~\cite{thz_Akyildiz2022May,nearFieldModel_Elbir2014Sep,elbir2023Feb_nearFieldBS_NBA_OMP}. For ULA, we have $G = (N_\mathrm{T}-1)\frac{c_0}{2f_c}$, for which the Fraunhofer distance becomes $R \approx \frac{ N_\mathrm{T}^2 c_0}{2f_c} $ for large $N_\mathrm{T}$.

	Compared to its far-field counterpart, the near-field model involves additional range parameter. Define $r$ as the distance between the user and the array center at the BS. Then, the $i$th element of the steering vectors for ULA in near- and far-field are given  by~\cite{thz_nearField_BeamSplit_Cui2021Sep}
	\begin{align}
	\label{nfModel}
	[\mathbf{a}_\mathrm{NF} (\vartheta,r)]_i &= e^{-j2\pi \frac{f_m}{c_0}  (r^{(i)} -r) } \\
	[\mathbf{a}_\mathrm{FF} (\vartheta)]_i &= e^{j2\pi \bar{d}\frac{f_m}{c_0} (i-1) \vartheta  },
	\end{align}
	where $r^{(i)}$ is the distance between the user and the $i$th BS antenna as
	\begin{align}
	\label{range1}
	r^{(i)} &= r \sqrt{1 + \left(\frac{(i-1) \bar{d}}{r}\right)^2  - \frac{2(i-1) \bar{d} \vartheta}{r}  }.
	\end{align}
	Applying Taylor series expansion to the right hand side of (\ref{range1})~\cite{nearFieldModel_Elbir2014Sep,nearFieldDOA_Jiang2013Apr}, we get
	\begin{align}
	\label{range2}
	r^{(i)} = r - (i-1)\bar{d} \vartheta + \frac{\left((i-1)\bar{d} \cos \tilde{\vartheta}\right)^2}{2 r},
	\end{align}
	where the last term tends to $0$ in the far-field scenario, i.e., $	e^{-j2\pi \frac{f_m}{c_0} \left((r- (i-1)\bar{d}\vartheta)  -r\right)} = e^{j \pi \frac{f_m}{f_c} (i-1)\vartheta} $.		Now, we calculate beam-split in near-field. Using (\ref{range2}) and $\bar{d} = \frac{c_0}{2f_c}$, we rewrite (\ref{nfModel}) as
	\begin{align}
	\label{nf_sv}
	[\mathbf{a}_\mathrm{NF} (\vartheta,r)]_i &= e^{-j2\pi \frac{f_m}{c_0}  ( -(i-1)\bar{d}\vartheta    + \frac{\left((i-1)\bar{d} \cos \tilde{\vartheta}\right)^2}{2 r})  } \nonumber\\
	& = e^{j(\pi  \frac{f_m}{f_c} (i-1)(\vartheta   - \frac{ c_0(i-1)\cos^2\tilde{\vartheta}}{4 f_c r})   ) }\nonumber\\
	& = e^{j\pi(i-1)\theta_{r,m,i}  },
	\end{align}
	where  $\theta_{r,m,i}$ denotes the distance-dependent near-field spatial channel direction as
	\begin{align}
	\theta_{r,m,i} =   \frac{f_m}{f_c} \vartheta   - \frac{f_m c_0 (i-1)\cos^2 \tilde{\vartheta}}{4f_c^2r}.
	\end{align}
	Then, the near-field beam-split is defined as
	\begin{align}
	\Delta_{r,i}[m] &= \theta_{r,m,i} - \vartheta \nonumber\\
	&= (\frac{f_m}{f_c} - 1)\vartheta  - \frac{f_m c_0 (i-1)\cos^2 \tilde{\vartheta}}{4f_c^2r}.
	\end{align}
	Notice that the near-field beam-split also depends on the antenna index $i$ with an additional distance-dependent term $- \frac{f_m c_0 (i-1)\cos^2 \tilde{\vartheta}}{4f_c^2r}$	 while it converges to far-field beam-split i.e., $\Delta_{r,i}[m] \approx \Delta [m]$ as the user distance increases since $\lim_{r \rightarrow \inf } \frac{f_m c_0 (i-1)\cos^2 \tilde{\vartheta}}{4f_c^2r} = 0$.

	%
	%

	\section{SBFL for THz Channel Estimation}
	\label{sec:SBFL}
	Since the THz system involves the UM number of antennas, the channel estimation stage may be computationally complex, for which the data-driven techniques, such as ML and FL can be employed to ease the computational burden. Hence, we propose learning-based approach, i.e., SBFL, for THz channel estimation. While traditional ML method can also be considered, FL is communication-efficient since it does not involve dataset transmission~\cite{elbir_FL_PHY_Elbir2021Nov}. Therefore, we consider an FL-based training scheme to provide a communication-efficient solution for THz channel estimation.
	
	Denote $\boldsymbol{\xi}\in\mathbb{R}^Q$  as the vector of real-valued model parameters, and define $\mathcal{D}_k$ as the dataset of the $k$th user. By training the model $\boldsymbol{\xi}$ on $\mathcal{D}_k$  allows us to construct the nonlinear function $f(\boldsymbol{\xi})$  as $\mathcal{Y}_k^{(i)} = f(\boldsymbol{\xi}) \mathcal{X}_k^{(i)}$, $i = 1,\dots, \textsf{D}_k$. Here,  $\textsf{D}_k = |\mathcal{D}_k|$ stands for the number of samples for the $k$th  dataset. Also, $\mathcal{X}_k^{(i)}$ and $\mathcal{Y}_k^{(i)}$, respectively,  denote the input and output data for the $i$th sample of the $k$th dataset. Thus, we have $\mathcal{D}_k^{(i)} = (\mathcal{X}_k^{(i)},\mathcal{Y}_k^{(i)})$.

	\subsection{Training}
	To begin with, we construct the input-output data used for training. We define the input data by $\mathcal{X}_k\in \mathbb{R}^{P\times 3}$, which involves the real, imaginary parts and angle of the received pilot signal $\mathbf{y}_k[m]$. Note that we use a \textit{three-channel} input data  for enhanced feature extraction performance~\cite{elbir_FL_PHY_Elbir2021Nov,elbir2020_FL_CE}. As a result, we have  $[\mathcal{X}_k]_1 = \operatorname{Re}\{\mathbf{y}_k[m]\}$, $[\mathcal{X}_k]_2 = \operatorname{Im}\{\mathbf{y}_k[m]\}$ and $[\mathcal{X}_k]_3 = \angle\{\mathbf{y}_k[m]\}$. In addition, the channel estimate obtained via SBCE, i.e., $\hat{\mathbf{h}}_k[m]$, is used to construct the output data. Thus, we define the output  as	$\mathcal{Y}_k = \left[\operatorname{Re}\{\hat{\mathbf{h}}_k[m]\}^\textsf{T},\operatorname{Im}\{\hat{\mathbf{h}}_k[m]\}^\textsf{T}\right]^\textsf{T}\in \mathbb{R}^{2N_\mathrm{T}}$.
	
	Now, we state the FL-based training model as 
	\begin{align}
	\label{flTraining}
	\minimize_{\boldsymbol{\xi}} \hspace{5pt} &\frac{1}{K}\sum_{k=1}^K \mathcal{L}_k(\boldsymbol{\xi}) \nonumber \\
	\subjectto \hspace{3pt} &f(\mathcal{X}_k^{(i)}| \boldsymbol{\xi}) = \mathcal{Y}_k^{(i)},
	\end{align}
	where $i = 1,\dots, \textsf{D}_k$ and $k = 1\dots, K$. In (\ref{flTraining}), $\mathcal{L}_k (\boldsymbol{\xi}) = \frac{1}{\textsf{D}_k} \sum_{i=1}^{\textsf{D}_k} ||f(\mathcal{X}_k^{(i)}| \boldsymbol{\xi}) - \mathcal{Y}_k^{(i)} ||_{\mathcal{F}}^2$ denotes to the loss function at the $k$th user and the  wireless channel estimate at the $k$th user is given by $f(\mathcal{X}_k^{(i)}| \boldsymbol{\xi})$ when the input is $\mathcal{X}_k^{(i)}$. The FL problem in (\ref{flTraining}) is effectively solved via stochastic gradient descent algorithms, for which the learning model parameters $\boldsymbol{\xi}$ is iteratively computed by each user, and then aggregated at the cloud server via BS\footnote{In a general setting, the computational resources of all users are assumed to be satisfactory. Otherwise, hybrid federated/centralized learning strategies can be followed~\cite{elbir_HFCL_Elbir2022Jun}. Then, the dataset of the users who are not able to compute the model parameters are sent to the BS, which performs computation on behalf of them.}. Therefore, we consider the model update rule for the $j$th iteration as
	\begin{align}
	\boldsymbol{\xi}_{j+1} = \boldsymbol{\xi}_j - \varepsilon \frac{1}{K} \sum_{k=1}^{K}\boldsymbol{\beta}_k(\boldsymbol{\xi}_j),
	\end{align} 
	for $j = 1,\dots, J$. Here,  $\boldsymbol{\beta}_k(\boldsymbol{\xi}_j) = \nabla \mathcal{L}_k(\boldsymbol{\xi}_j)\in\mathbb{R}^Q$ is composed of the gradients of the model parameter $\boldsymbol{\xi}_j$ and $\varepsilon$ is the learning rate. 
	

	\subsection{Communication Overhead}
	\label{sec:complexity_overhead}
	Similar to the previous works, the communication overhead is defined as the amount of transmitted data during model training~\cite{elbir2020_FL_CE,fl_By_Google,fl_spm_federatedLearning}. In particular, the overhead of CL mainly due to the dataset transmission of $K$ users whereas the overhead of FL involves the model parameter transmission  of $K$ users for $J$ iterations. As a result, we define the overhead of CL and FL as  $\mathcal{T}_\mathrm{FL} = 2 Q JK$ and $\mathcal{T}_\mathrm{CL} = \sum_{k = 1}^{K}\textsf{D}_k(3P + 2N_\mathrm{T})$, respectively.

	\begin{figure}[t]
		\centering
		{\includegraphics[draft=false,width=\columnwidth]{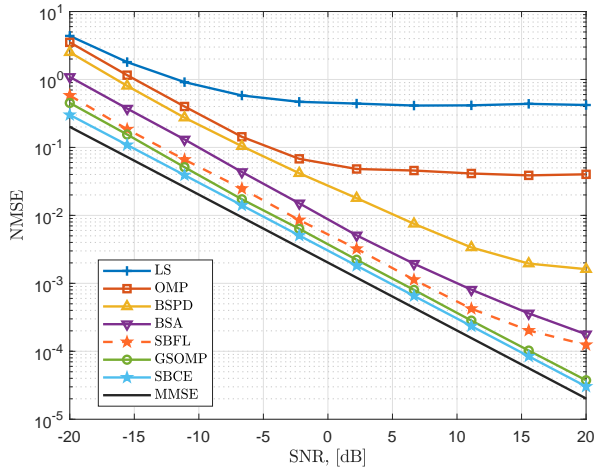} } 
		\caption{Channel estimation NMSE versus SNR. 
		}
		\label{fig_NMSE_SNR}
	\end{figure}

	\section{Numerical Experiments}
	\label{sec:Sim}
	In this section, the performance of  our  SBCE technique is evaluated and compared with the state-of-the-art techniques such as LS and MMSE in (\ref{ls_MMSE}), GSOMP~\cite{dovelos_THz_CE_channelEstThz2}, OMP, BSPD~\cite{thz_channelEst_beamsplitPatternDetection_L_Dai} and BSA~\cite{elbir2022Jul_THz_CE_FL}. Note that MMSE is selected as benchmark since it relies on the channel covariance matrix at each subcarrier, hence it presents beam-split-free performance. The NMSE  is computed as $\mathrm{NMSE} = \frac{1}{J_\mathrm{T}KM}\sum_{i=1,k=1,m=1}^{J_\mathrm{T},K,M}\frac{|| \mathbf{h}_k[m]- \hat{\mathbf{h}}_k^{(i)}[m] ||_2^2 }{ || \mathbf{h}_k[m]||_2^2 }$, where $J_\mathrm{T}=100$ is the number of trials. The simulation parameters for the THz channel scenario are $f_c=300$ GHz, $B=30$ GHz, $M=128$, $N_\mathrm{T} = 256$, $N_\mathrm{RF}=P=32$, $L=1$, $K=8$ and $\tilde{\vartheta}_{k}\in \mathrm{unif}[-\frac{\pi}{2},\frac{\pi}{2}]$~\cite{elbir2021JointRadarComm,thz_channelEst_beamsplitPatternDetection_L_Dai,ummimoTareqOverview}.  $\mathbf{D}$ is formed  with $N = 8N_\mathrm{T}$, and $[\mathbf{B}]_{i,j} = \frac{1}{\sqrt{N_\mathrm{T}}}e^{j\tilde{\varphi}}$, where $\tilde{\varphi} \sim \mathrm{unif}(-1,1)$. The proposed SBCE algorithm is initialized with $\boldsymbol{\sigma}^{(0)} \hspace{-2pt}=\hspace{-2pt} \mathbf{1}_{N}$, $\mathbf{C}_k^{(0)}[m] \hspace{-2pt}= \hspace{-2pt}\mathbf{I}_{N_\mathrm{T}}$, ${\mu^2}^{(0)} \hspace{-2pt}=\hspace{-2pt} 0$, and run as presented in Algorithm~\ref{alg:SBCE}. The termination parameter is selected as  $\epsilon_\mathrm{TH} = 0.001$. It is observed that the SBCE algorithm converges in approximately $T=50$ iterations. 
	
	For SBFL, the FL toolbox in MATLAB is employed for data generation and training. The learning model is a CNN with $12$ layers and $Q=603,648$ learnable parameters~\cite{elbir2020_FL_CE,elbir2022Jul_THz_CE_FL}. Specifically, the first layer has the size of $P\times 3$ for input. The $\{2,4,6,8\}$th layers are convolutional layers, and they are constructed with $128$ kernels of size $3\times 3$. Each convolutional layer is followed by a normalization layer in the $\{3,5,7,9\}$th layers. There is a fully-connected layer in the $10$th layer of size $1024$, which is followed by  a dropout layer. The last  $12$th layer is the output layer of size $2N_\mathrm{T}\times 1$. During data generation, $100$ channel realizations are generated. In order to provide robust performance against the imperfections in the receive data, we added AWGN on the input received pilot data for three signal-to-noise-ratio (SNR) levels, i.e., $\text{SNR} = \{15,20,25\}$ dB and  $ 50$ noisy realizations~\cite{elbir2020_FL_CE}. The CNN is trained for $J=100$ with the rate of $\varepsilon = 0.001$.

	\begin{figure}[t]
		\centering
		{\includegraphics[draft=false,width=\columnwidth]{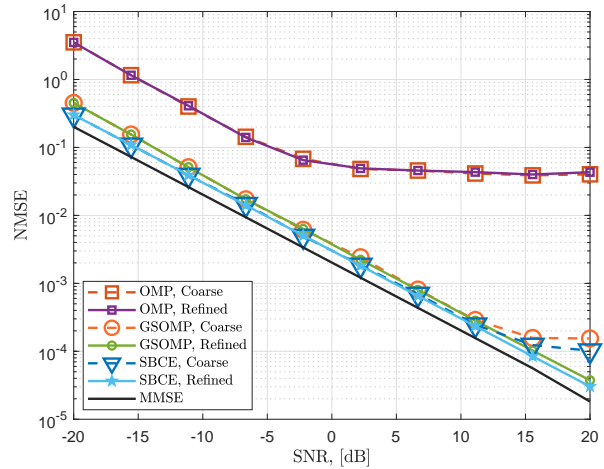} } 
		\caption{Channel estimation NMSE versus SNR for coarse and refined dictionaries. 
		}
		\label{fig_NMSE_SNR_refined}
	\end{figure}

	Fig.~\ref{fig_NMSE_SNR} shows the NMSE performance of the competing algorithms with respect to SNR. We observe that the proposed SBCE approach attains better NMSE as compared to other algorithmic-based approaches, i.e., GSOMP, BSA and BSPD. The direct application of both OMP and LS yields poor NMSE since they do not involve any specific mechanism to mitigate beam-split. While BSPD and BSA are beam-split-aware techniques, they suffer from inaccurate support detection, which leads to precision loss as SNR increases. \textcolor{black}{Although GSOMP employs SD dictionary matrices with TDs, SBCE  exhibits lower NMSE than that of GSOMP. This is mainly due to two reasons that SBCE performs better support recovery than GSOMP, which is an OMP-based algorithm~\cite{sbl_betterthanOMP_Srivastava2018Dec,sbl_varX_Wu2015Dec,sbl_versus_CS_Tan2009Jan}, and GSOMP does not include refined direction estimation stage.} {\color{black}In order to assess impact of refined direction estimation, Fig.~\ref{fig_NMSE_SNR_refined} shows the channel estimation performance with coarse and refined direction estimates. During simulations the same coarse dictionary is employed for OMP, GSOMP and SBCE. Then, refined direction estimation method in Sec.~\ref{sec:Refine} is run based on the estimate obtained from each algorithm. We can see that no significant improvement is obtained for OMP algorithm since it does not take into account the impact of beam-split. However, improved NMSE is achieved, especially in high SNR, for GSOMP and SBCE. Again, we obtain similar observations as in~Fig.~\ref{fig_NMSE_SNR}, that is, SBCE has better performance than that of GSOMP in both coarse and refined direction scenario thanks to its accurate support recovery.  }

\begin{figure}[t]
		\centering
		{\includegraphics[draft=false,width=\columnwidth]{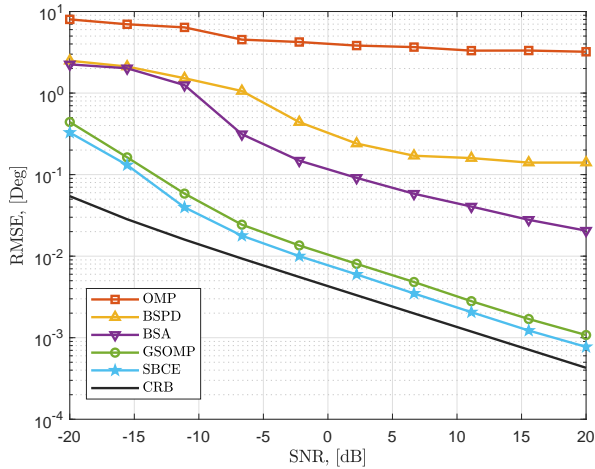} } 
		\caption{ Physical channel direction RMSE versus SNR. 
		}
		\label{fig_RMSE_SNR}
	\end{figure}

	In order to demonstrate the physical direction estimation performance, RMSE of the estimated path directions is presented in Fig.~\ref{fig_RMSE_SNR}. While the proposed SBCE approach follows the CRB closely, the remaining methods suffer from grid error and yield approximately $0.1^\circ$ RMSE. 
	
	Comparing SBCE and SBFL, we observe that the latter has slight performance loss as well as it suffers from low NMSE in high SNR. This is because of the neural network losing precision due to decentralize learning and noisy dataset to prove robustness against imperfect data. Nevertheless, the main advantage of FL is the communication-efficiency, i.e., model training with less data/model transmission overhead. According to the analysis in Sec.~\ref{sec:complexity_overhead}, the communication overhead of CL is computed as $\mathcal{T}_\mathrm{CL} = 8\cdot\textsf{D}_k\cdot (3\cdot32 + 2\cdot128) = 10.8\times 10^{9}	$, where the number of input-output tuples in the dataset is $\textsf{D}_k = 3M VG = 3\cdot 128\cdot 100\cdot 100 = 3.07\times 10^{6} $. On the other hand, the communication overhead of FL is $\mathcal{T}_\mathrm{FL} = 2\cdot 603,648\cdot 100\cdot 8 = 0.96\times 10^{9}$, which exhibits approximately $11$ times lower than that of CL. Thus, the proposed SBFL approach is an efficient tool especially  for THz-related datasets, which usually involve large number of antennas.

\begin{figure}[t]
		\centering
		{\includegraphics[draft=false,width=\columnwidth]{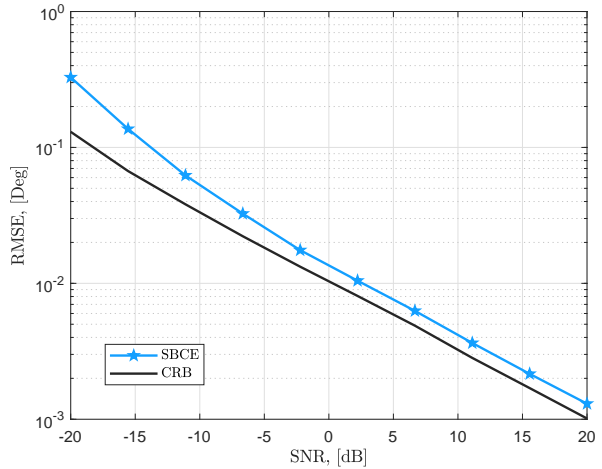} } 
		\caption{Beam-split estimation RMSE versus SNR. 
		}
		\label{fig_RMSE_BS_SNR}
	\end{figure}

	In Fig.~\ref{fig_RMSE_BS_SNR}, beam-split estimation RMSE is shown with respect to SNR. As it is seen, the proposed approach effectively estimates the beam-split and follows the statistical lower bound CRB, which is derived in Appendix~\ref{appendixCRB}, with a slight performance gap (e.g., $\sim 0.01^\circ $ in $0$ dB). The channel estimation techniques, i.e., GSOMP, BSPD and BSA, however, do not have the capability to obtain the beam-split.

\begin{figure}[t]
		\centering
		{\includegraphics[draft=false,width=\columnwidth]{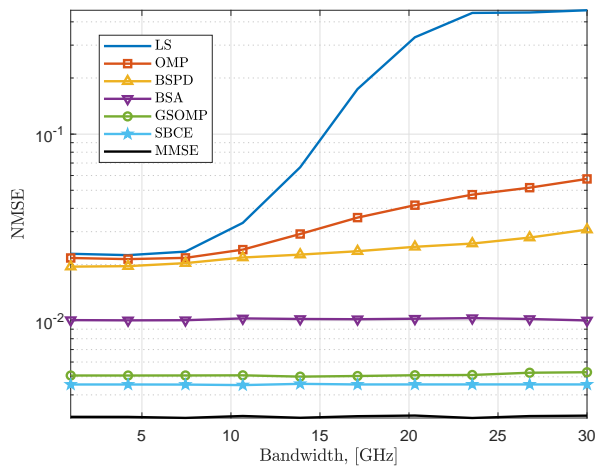} } 
		\caption{Channel estimation NMSE against bandwidth. 
		}
		\label{fig_NMSE_bandwidth}
	\end{figure}

	Fig.~\ref{fig_NMSE_bandwidth} shows the channel estimation NMSE against bandwidth for the interval of $[1,30]$ GHz at $300$ GHz carrier frequency. We can see that all of the algorithms provide approximately $0.02$ NMSE for the bandwidth $B<7$ GHz. However, the performance of LS and OMP degrades as the bandwidth widens since these method do not involve beam-split mitigation. Also, BSPD yields a slight NMSE loss at wide bandwidth while BSA has robust performance. On the other hand, the proposed SBCE method enjoys robustness against the increase of the bandwidth.


\begin{figure}[t]
		\centering
		\subfloat[]{\includegraphics[draft=false,width=\columnwidth]{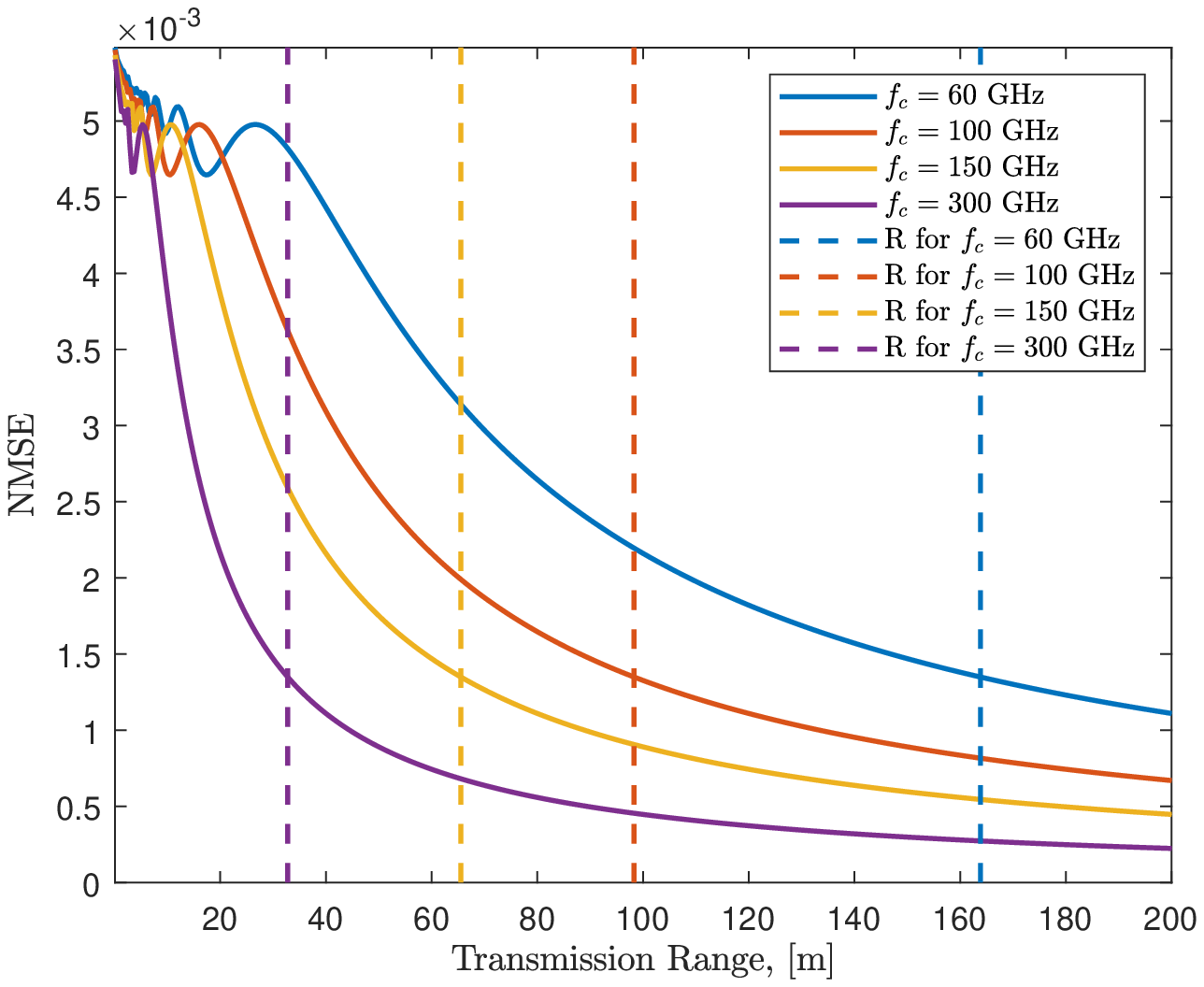} }\\
		\subfloat[]{\includegraphics[draft=false,width=\columnwidth]{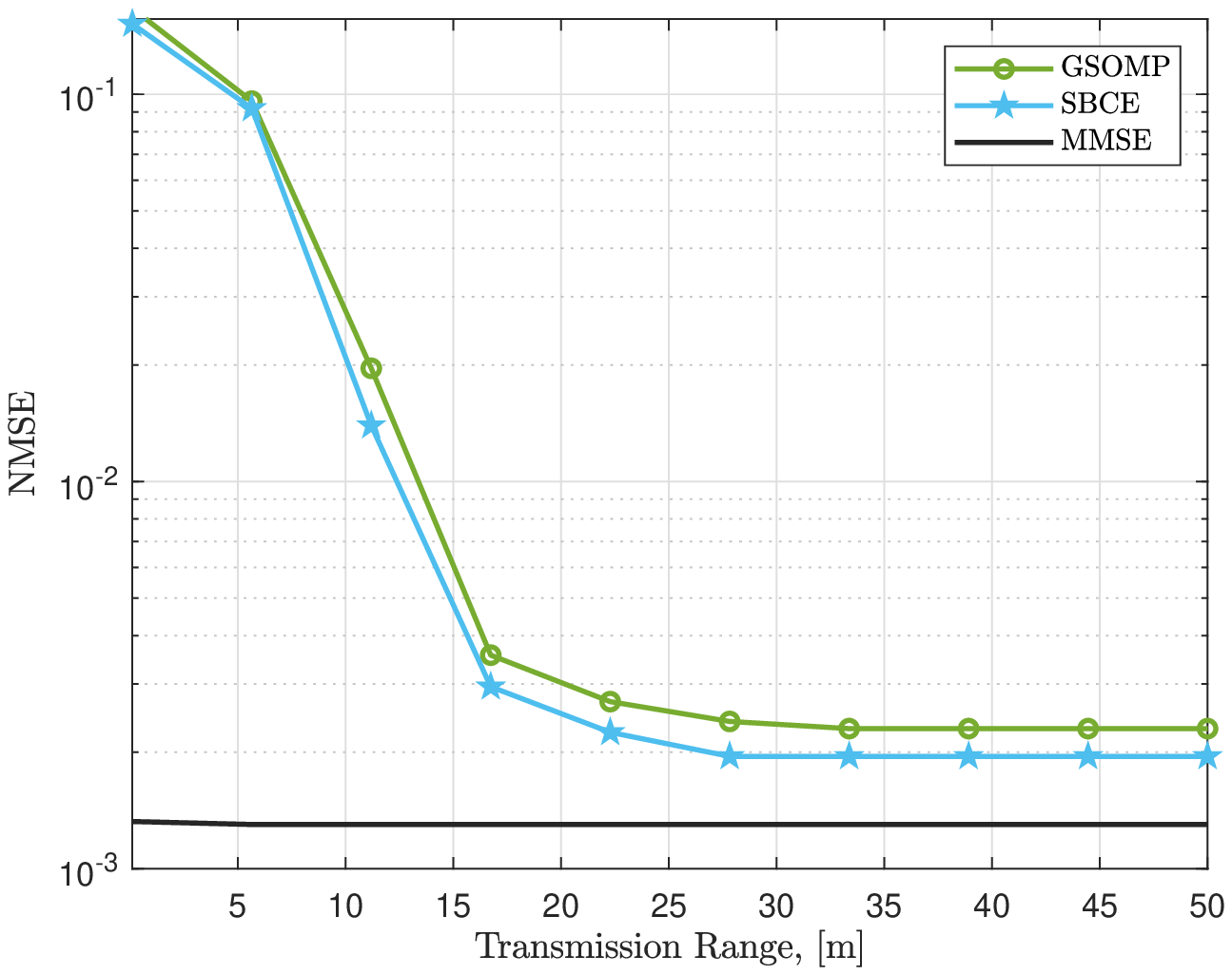} } 
		\caption{(a) NMSE between near- and far-field steering vectors, and (b) channel estimation NMSE for the usage of far-field model with respect to transmission range. 
		}
		\label{fig_NMSE_Range}
	\end{figure}

	Finally, we investigate the deviation from far-field model in terms of transmission range and frequency. Fig.~\ref{fig_NMSE_Range}(a) shows the normalized difference between near- and far-field steering vectors, i.e., $\frac{\| \mathbf{a}_\mathrm{NF} - \mathbf{a}_\mathrm{FF}\|_2^2}{\|\mathbf{a}_\mathrm{FF}\|_2^2}$, with respect to transmission range for various frequencies and corresponding Fraunhofer distances. As it is seen, the NMSE curve crosses the Fraunhofer distance approximately at $0.0013$ for all frequencies at different ranges. Specifically, NMSE crosses the Fraunhofer distance at $32$ m for $f_c = 300$ GHz, which shows that the near-field model should be considered for $r< 32$ m when $N_\mathrm{T} = 256$. Note that this distance is smaller if rectangular arrays are used. For instance, a $16\times 16$ URA has the array aperture of  $G = 16\sqrt{2}\bar{d}$ leading to $R =0.256$ m, which yields much safer use of far-field model in shorter distance. Fig.~\ref{fig_NMSE_Range}(b) illustrates the channel estimation NMSE with respect to transmission range when far-field model is used in GSOMP and SBCE. We observe that relatively large error for $r < R$ due to mismatch between the far- and near-field steering vectors. Although fixing the direction information may yield less error~\cite{dovelos_THz_CE_channelEstThz2}, which only due to range-mismatch, the mismatch in the range also causes inaccurate direction estimates.

	\section{Conclusions}
	\label{sec:Conc}
	In this work, we proposed an SBL-based approach for joint THz channel and beam-split estimation. The proposed method is based on modeling the beam-split as an array perturbation. We have shown that the proposed SBCE approach effectively estimates THz channel as compared to existing methods without requiring additional hardware components such as TD networks. The SBCE technique is also capable of effectively estimating the beam-split. This is particularly helpful to design the hybrid beamformers more easily with the knowledge of beam-split. Also, a model-free technique, SBFL, is introduced to realize the problem from ML perspective for computational- and communication-efficiency.  Furthermore, we examined the near-field considerations of the THz channel and evaluate the performance of SBCE with far-field model. The estimation of the near-field THz channel is one particular problem that we reserve to study for future work.

	\appendices
	\section{Proof of Lemma~\ref{lemmaArrayGain}}
	\label{appendixAG}
	The array gain varies across the whole bandwidth as
	\begin{align}
	\label{arrayGain2}
	A_G({\vartheta},\theta_{m}) = \frac{|\mathbf{a}^\textsf{H}(\vartheta)  \mathbf{a}'(\theta_m)|^2}{N_\mathrm{T}^2}.
	\end{align}
	By using (\ref{steeringVec}), (\ref{arrayGain2}) is rewritten  as
	\begin{align}
	A_G({\vartheta},{\theta_m})&= \frac{1}{N_\mathrm{T}^2} \left| \sum_{n_1 =1}^{N_\mathrm{T}}  \sum_{n_2=1}^{N_\mathrm{T}} e^{-\mathrm{j} \pi  \left( (n_1-1){\theta_m} - (n_2-1)\frac{\lambda_c\vartheta}{\lambda_m}\right)    }   \right| ^2 \nonumber \\
	&=\frac{1}{N_\mathrm{T}^2} \left|\sum_{n = 0}^{N_\mathrm{T}-1} e^{-\mathrm{j}2\pi n d \left( \frac{\theta_m}{\lambda_c} - \frac{\vartheta}{\lambda_m}  \right)     }   \right|^2 \nonumber\\
	&=  \frac{1}{N_\mathrm{T}^2} \left|\sum_{n = 0}^{N_\mathrm{T}-1} e^{-\mathrm{j}2\pi n d \frac{(f_c \theta_m - f_m\vartheta) }{c_0}     }   \right|^2 \nonumber\\
	&= \frac{1}{N_\mathrm{T}^2} \left| \frac{1 - e^{-\mathrm{j}2\pi N_\mathrm{T}d \frac{(f_c \theta_m - f_m\vartheta)}{c_0}    }}{1 - e^{-\mathrm{j}2\pi d\frac{(f_c \theta_m - f_m\vartheta)}{c_0}  }}   \right|^2  \nonumber\\
	&= \frac{1}{N_\mathrm{T}^2}\left| \frac{\sin (\pi N_\mathrm{T}\mu_m )}{\sin (\pi \gamma_m )}    \right|^2 \nonumber\\
	& = \left||\xi( d\frac{(f_c \theta_m - f_m\vartheta)}{c_0})\right|^2. \label{arrayGain}
	\end{align}
	The array gain in (\ref{arrayGain}) implies that most of the power is focused only on a small portion of the beamspace due to the power-focusing capability of $\xi(a)$, which substantially reduces across the subcarriers as $|f_m - f_c|$ increases. Furthermore, $|\xi( \mu_m )|^2$ gives peak when $\mu_m = 0$, i.e.,  $f_c \theta_m - f_m\vartheta= 0$. Thus, we have $ \theta_m = \eta_m \vartheta$, which completes the proof.  \qed
	
	{\color{black}
		\section{Derivation  of (\ref{y_Px_posterior1})}
		\label{appendix_Posterior1}
		In order to obtain (\ref{y_Px_posterior1}), we utilize the Bayes rule to obtain the posterior density of $\mathbf{x}$ as $p(\mathbf{x}| \mathbf{y}  ) \sim  \mathcal{CN} (\mathbf{z}, \boldsymbol{\Pi})$ where the mean and the variance are
		\begin{align}
		\label{mean_z}
		\mathbf{z} = \mu^{-2}\boldsymbol{\Pi} {\mathbf{P}'}^{\textsf{H}}\mathbf{y},
		\end{align} 
		and 
		\begin{align}
		\label{variance_z}
		\boldsymbol{\Pi} &=  \left[ \boldsymbol{\Sigma}^{-1} + \mu^{-2} {\mathbf{P}'}^\textsf{H} {\mathbf{P}'} \right]^{-1} ,
		\end{align} respectively~\cite{sbl_doa_Liu2012Sep,sbl_outlier_Dai2017Nov}. Now, $\mathbb{E}\{  \|\mathbf{y} - \mathbf{P}' \mathbf{x} \|_2^2  \}$ is rewritten  as
		\begin{align}
		\mathbb{E}\{  \|\mathbf{y} - \mathbf{P}' \mathbf{x} \|_2^2  \} 
		=  \mathbb{E}\{  \|\mathbf{y} - \mathbf{P}' \mathbf{z} \|_2^2  + \| \mathbf{P}' (\mathbf{x} - \mathbf{z})   \|_2^2 \},\label{e_px_z}
		\end{align}
		where 
		\begin{align}
		\mathbb{E}\{  \| &\mathbf{P}' (\mathbf{x} - \mathbf{z})   \|_2^2 \}	 =  \mathbb{E}\{   \mathrm{Tr}\{ (\mathbf{x} - \mathbf{z})^\textsf{H} {\mathbf{P}'}^\textsf{H} \mathbf{P}' (\mathbf{x} - \mathbf{z})       \}\} \nonumber \\
		&= \mathbb{E}\{   \mathrm{Tr}\{ (\mathbf{x} - \mathbf{z})^\textsf{H} {\mathbf{P}'}^\textsf{H} \mathbf{P}' (\mathbf{x} - \mathbf{z})       \}\} \nonumber \\
		&= \mathbb{E}\{   \mathrm{Tr}\{ {\mathbf{P}'} (\mathbf{x} - \mathbf{z})  (\mathbf{x} - \mathbf{z})^\textsf{H} {\mathbf{P}'}^\textsf{H}      \}\}, \nonumber \\
		&=    \mathrm{Tr}\{ {\mathbf{P}'} \mathbb{E}\{(\mathbf{x} - \mathbf{z})  (\mathbf{x} - \mathbf{z})^\textsf{H}  \}{\mathbf{P}'}^\textsf{H}     \}. \label{trace__x_z}
		\end{align}
		In order to compute  $ \mathbb{E} \{ (\mathbf{x}- \mathbf{z}) (\mathbf{x} - \mathbf{z})^\textsf{H}  \}$, we first obtain $\mathbb{E} \{ \mathbf{z} \mathbf{z}^\textsf{H} \}$ by utilizing (\ref{mean_z}) as
		\begin{align}
		\mathbf{z} = \mu^{-2}\boldsymbol{\Pi} {\mathbf{P}'}^{\textsf{H}}\mathbf{y},
		\end{align}
		where $\boldsymbol{\Pi}$ from (\ref{variance_z}) can be written via applying matrix inversion lemma as
		\begin{align}
		\label{Pi_matrixInversionLemma}
		\boldsymbol{\Pi} = \boldsymbol{\Sigma} -  \boldsymbol{\Sigma}{\mathbf{P}'}^\textsf{H} \boldsymbol{\Pi}_\mathbf{y}^{-1}\mathbf{P}' \boldsymbol{\Sigma},
		\end{align}
		for which we have
		\begin{align}
		\mu^{-2}\boldsymbol{\Pi} {\mathbf{P}'}^\textsf{H}\mathbf{y} &= \mu^{-2}\boldsymbol{\Sigma}{\mathbf{P}'}^\textsf{H}\mathbf{y} - \mu^{-2} \boldsymbol{\Sigma}{\mathbf{P}'}^\textsf{H} \boldsymbol{\Pi}_\mathbf{y}^{-1}\mathbf{P}' \boldsymbol{\Sigma}{\mathbf{P}'}^\textsf{H}\mathbf{y} \nonumber \\
		& = \mu^{-2}\boldsymbol{\Sigma}{\mathbf{P}'}^\textsf{H} \underbrace{\left[ \mathbf{I}_P -  \boldsymbol{\Pi}_\mathbf{y}^{-1}\mathbf{P}' \boldsymbol{\Sigma}{\mathbf{P}'}^\textsf{H} \right]}_{  \mu^{2} \boldsymbol{\Pi}_\mathbf{y}^{-1}   } \mathbf{y} \nonumber \\
		& = \boldsymbol{\Sigma}{\mathbf{P}'}^\textsf{H}  \boldsymbol{\Pi}_\mathbf{y}^{-1}\mathbf{y} = \mathbf{z},\label{obtain_z}
		\end{align}
		for which the second line in the right hand side involves 
		\begin{align}
		\boldsymbol{\Pi}_\mathbf{y} &= \mathbf{P}' \boldsymbol{\Sigma}{\mathbf{P}'}^\textsf{H} + \mu^2 \mathbf{I}_P \nonumber \\
		\mathbf{I}_P& =  \boldsymbol{\Pi}_\mathbf{y}^{-1}\mathbf{P}' \boldsymbol{\Sigma}{\mathbf{P}'}^\textsf{H} + \mu^2\boldsymbol{\Pi}_\mathbf{y}^{-1}
		\mu^{2} \nonumber \\
		\boldsymbol{\Pi}_\mathbf{y}^{-1} & = \mathbf{I}_P -  \boldsymbol{\Pi}_\mathbf{y}^{-1}\mathbf{P}' \boldsymbol{\Sigma}{\mathbf{P}'}^\textsf{H}.
		\end{align}
		Then, we have  $ \mathbb{E} \{ (\mathbf{x}- \mathbf{z}) (\mathbf{x} - \mathbf{z})^\textsf{H}  \} =\mathbb{E} \{ \mathbf{x} \mathbf{x}^\textsf{H} \} + \mathbb{E} \{ \mathbf{z} \mathbf{z}^\textsf{H} \} $ due to their independence, and combining with (\ref{obtain_z}) yields
		\begin{align}
		\mathbb{E} \{ \mathbf{x} \mathbf{x}^\textsf{H} \} + \mathbb{E} \{ \mathbf{z} \mathbf{z}^\textsf{H} \} &= \boldsymbol{\Sigma}  + \boldsymbol{\Sigma}{\mathbf{P}'}^\textsf{H}  \boldsymbol{\Pi}_\mathbf{y}^{-1} \mathbb{E} \{\mathbf{yy}^\textsf{H}  \}   \boldsymbol{\Pi}_\mathbf{y}^{-1} {\mathbf{P}'} \boldsymbol{\Sigma}^\textsf{H} \nonumber \\
		&=\boldsymbol{\Sigma}  + \boldsymbol{\Sigma}{\mathbf{P}'}^\textsf{H}     \boldsymbol{\Pi}_\mathbf{y}^{-1} \boldsymbol{\Pi}_\mathbf{y}\boldsymbol{\Pi}_\mathbf{y}^{-1} {\mathbf{P}'} \boldsymbol{\Sigma}^\textsf{H}\nonumber \\
		&=\boldsymbol{\Sigma}  + \boldsymbol{\Sigma}{\mathbf{P}'}^\textsf{H}     \boldsymbol{\Pi}_\mathbf{y}^{-1} {\mathbf{P}'} \boldsymbol{\Sigma}^\textsf{H} \nonumber \\
		& = \boldsymbol{\Pi}, \label{Pi_final}
		\end{align}
		due to (\ref{Pi_matrixInversionLemma}). Finally, combining (\ref{Pi_final}) and (\ref{trace__x_z}),  we get 
		\begin{align}
		\mathrm{Tr}\{ {\mathbf{P}'} \mathbb{E}\{(\mathbf{x} - \mathbf{z})  (\mathbf{x} - \mathbf{z})^\textsf{H}  \}{\mathbf{P}'}^\textsf{H}     \} = \mathrm{Tr}\{\mathbf{P}' \boldsymbol{\Pi} {\mathbf{P}'}^\textsf{H} \}. 
		\end{align}
		Thus,  (\ref{e_px_z}) becomes
		\begin{align}
		\mathbb{E}\{  \|\mathbf{y} - \mathbf{P}' \mathbf{x} \|_2^2  \}  =  ||\mathbf{y} - {\mathbf{P}'} \mathbf{z}  ||_2^2 + \mathrm{Tr}\{\mathbf{P}' \boldsymbol{\Pi} {\mathbf{P}'}^\textsf{H} \}.
		\end{align}

	}

	\section{Cram{\'e}r-Rao Lower Bound}
	\label{appendixCRB}
	In order to derive the CRB, we assume a single user case, i.e., $K=1$ without loss of generality. The CRB expressions can easily be extended for the $K>1$ case by using the decoupledness of the unknown parameters of multiple users. Define the unknown parameter vector as the physical directions and beam-splits corresponding to $L$ paths. Hence, we have a $2L\times 1$ unknown vector as 
	\begin{align}
	\label{unknownVec}
	\mathbf{v} = [\vartheta_1,\cdots,\vartheta_L, \Delta_1,\cdots,\Delta_L]^\textsf{T}\in \mathbb{C}^{2L}.
	\end{align}
	Consider the log-likelihood function of the joint pdf in (\ref{pdf_y}) with respect to $\mathbf{v}$ as
	\begin{align}
	\label{likelihood1}
	\mathcal{L}(\mathbf{v}) = \ln \{p(\mathbf{y}|\mathbf{v} )  \} 
	= -\ln |\boldsymbol{\Pi}_\mathbf{y} | - \mathrm{Tr} \{\boldsymbol{\Pi}_\mathbf{y}^{-1}\mathbf{R}_\mathbf{y} \}.
	\end{align}
	In order to obtain the CRB, we first need to find the Fisher information matrix (FIM), which measures the amount of information contained for the unknown variables. Let $\mathbf{FIM}\in \mathbb{R}^{2L\times 2L}$ be the FIM. Then, the $(i,j)$th entry of $\mathbf{FIM}$ is calculated as the second derivative of  $\mathcal{L}(\mathbf{v})$ as
	\begin{align}
	\label{fim}
	[\mathbf{FIM}]_{ij} = - \mathbb{E}\left\{ \frac{\partial^2 \mathcal{L}(\mathbf{v})}{\partial v_i \partial v_j}  \right\},
	\end{align}
	for $i,j = 1,\cdots,2L$. 	Calculate the first derivatives of $\mathcal{L}(\mathbf{v})$ by using
	\begin{align}
	\frac{\partial \ln \lbrace| \boldsymbol{\Pi}_\mathbf{y}|\rbrace }{\partial v_i} &= \mathrm{Tr}\left\lbrace   \boldsymbol{\Pi}_\mathbf{y}^{-1} \frac{\partial \boldsymbol{\Pi}_\mathbf{y}}{\partial v_i} \right\rbrace \nonumber \\
	\frac{\partial \boldsymbol{\Pi}_\mathbf{y}^{-1}}{\partial v_i} &= -  \boldsymbol{\Pi}_\mathbf{y}^{-1} \frac{\partial  \boldsymbol{\Pi}_\mathbf{y}}{\partial v_i}  \boldsymbol{\Pi}_\mathbf{y}^{-1},
	\end{align}
	we get the first derivative of $\mathcal{L}(\mathbf{v})$ as
	\begin{align}
	\frac{\partial \mathcal{L}(\mathbf{v}) }{\partial v_i} &=  - \mathrm{Tr}\left\lbrace   \boldsymbol{\Pi}_\mathbf{y}^{-1} \frac{\partial \boldsymbol{\Pi}_\mathbf{y}}{\partial v_i} \right\rbrace + \boldsymbol{\Pi}_\mathbf{y}^{-1} \frac{\partial  \boldsymbol{\Pi}_\mathbf{y}}{\partial v_i}  \boldsymbol{\Pi}_\mathbf{y}^{-1} \mathbf{R}_\mathbf{y} \nonumber \\
	&= \mathrm{Tr}\left\{ \left(  \boldsymbol{\Pi}_\mathbf{y}^{-1} \frac{\partial \boldsymbol{\Pi}_\mathbf{y}}{\partial v_i}\right) \left(\boldsymbol{\Pi}_\mathbf{y}^{-1} \mathbf{R}_\mathbf{y} - \mathbf{I}_{P}  \right)  \right\},
	\end{align}
	and the second derivative becomes
	\begin{align}
	\frac{\partial^2 \mathcal{L}(\mathbf{v}) }{\partial v_i \partial v_j} &= \mathrm{Tr}\Bigg\{ \frac{\partial}{\partial v_j}\left(  \boldsymbol{\Pi}_\mathbf{y}^{-1} \frac{\partial \boldsymbol{\Pi}_\mathbf{y}}{\partial v_i}\right) \left(\boldsymbol{\Pi}_\mathbf{y}^{-1} \mathbf{R}_\mathbf{y} - \mathbf{I}_{P}  \right) \nonumber \\
	& +  \left(  \boldsymbol{\Pi}_\mathbf{y}^{-1} \frac{\partial \boldsymbol{\Pi}_\mathbf{y}}{\partial v_i}\right) \frac{\partial}{\partial v_j}\left(\boldsymbol{\Pi}_\mathbf{y}^{-1} \mathbf{R}_\mathbf{y} - \mathbf{I}_{P}  \right)  \Bigg\} \nonumber \\
	&= \mathrm{Tr}\Bigg\{ \frac{\partial}{\partial v_j}\left(  \boldsymbol{\Pi}_\mathbf{y}^{-1} \frac{\partial \boldsymbol{\Pi}_\mathbf{y}}{\partial v_i}\right) \left(\boldsymbol{\Pi}_\mathbf{y}^{-1} \mathbf{R}_\mathbf{y} - \mathbf{I}_{P}  \right) \nonumber \\
	& +  \left(  \boldsymbol{\Pi}_\mathbf{y}^{-1} \frac{\partial \boldsymbol{\Pi}_\mathbf{y}}{\partial v_i}\right) \frac{\partial}{\partial v_j} \boldsymbol{\Pi}_\mathbf{y}^{-1} \frac{\partial  \boldsymbol{\Pi}_\mathbf{y}}{\partial v_j}  \boldsymbol{\Pi}_\mathbf{y}^{-1} \mathbf{R}_\mathbf{y} \Bigg\}.
	\end{align}
	Since $\mathbb{E}\{\mathbf{R}_\mathbf{y} \} = \boldsymbol{\Pi}_\mathbf{y}$, (\ref{fim}) is rewritten as 
	\begin{align}
	[\mathbf{FIM}]_{ij}\hspace{-3pt} = \hspace{-3pt} -\hspace{-1pt} \mathbb{E}\left\{\hspace{-3pt} \frac{\partial^2 \mathcal{L}(\mathbf{v})}{\partial v_i \partial v_j} \hspace{-3pt} \right\}\hspace{-3pt} =\hspace{-3pt} \mathrm{Tr}\hspace{-1pt} \Bigg\{\hspace{-3pt} \frac{\partial}{\partial v_i} \boldsymbol{\Pi}_\mathbf{y}^{-1} \frac{\partial  \boldsymbol{\Pi}_\mathbf{y}}{\partial v_j}  \boldsymbol{\Pi}_\mathbf{y}^{-1} \mathbf{R}_\mathbf{y} \hspace{-3pt} \Bigg\},
	\end{align}
	which is the general FIM expression for Gaussian signals. Then, by following the steps in~\cite{elbir2016Jul_Sparse_MutualCoupling,crbStoicaNehorai}, the CRB expressions corresponding to the $(i,j)$th entries of the FIM is  given by 
	\begin{align}
	\label{crb}
	\mathrm{CRB}_{ij} = \frac{1}{ \frac{2}{\mu^2}\mathrm{Tr}\{\mathbf{M} \mathbf{K}_{ij}   \}    },
	\end{align}
	where $\mathbf{M} = \overline{\boldsymbol{\Sigma}}^\textsf{H}{\mathbf{A}'}^\textsf{H} \boldsymbol{\Pi}_\mathbf{y}^{-1} \mathbf{A}' \overline{\boldsymbol{\Sigma}}  \in \mathbb{C}^{L\times L}$, where $\overline{\boldsymbol{\Sigma}} $ is an $L\times L$ matrix comprised of signal powers. $\mathbf{K}_{ij}\in \mathbb{C}^{L\times L}$ includes the derivative of the actual steering matrix $\mathbf{A}'$ with respect to the unknown parameters, and it is given by
	\begin{align}
	\mathbf{K}_{ij} = [\frac{\partial \mathbf{A}'}{\partial v_i }]^\textsf{H} (\mathbf{I}_{N_\mathrm{T}} - \mathbf{A}' {\mathbf{A}'}^\dagger   )[\frac{\partial \mathbf{A}'}{\partial v_j }].
	\end{align}
	In particular, the $i$th entry of the derivative of $\mathbf{a}'(\vartheta_l)$, i.e., the $l$th column of $\mathbf{A}'$ with respect to  the physical direction $\tilde{\vartheta}_l$ and beam-split $\Delta_l$ are respectively given by
	\begin{align}
	\frac{\partial [\mathbf{a}']_i}{\partial \tilde{\vartheta}_l } &= {j \pi (i - 1 ) \frac{f_m}{f_c}} \cos\tilde{\vartheta}_l [\mathbf{a}']_i, \\
	\frac{\partial [\mathbf{a}']_i}{\partial \Delta_l } &= {j \pi (i - 1 )} [\mathbf{a}']_i,
	\end{align}
	where we have via (\ref{steeringVec}) and (\ref{beamSplit}) that $[\mathbf{a}']_i = e^{j\pi (i-1)\frac{f_m}{f_c}\sin \tilde{\vartheta}_l} = e^{j\pi (i-1)(\Delta_l + \sin\tilde{\vartheta}_l)}$. 
	
	In case of near-field scenario, the range parameters are also added to the unknown vector defined in (\ref{unknownVec}), hence its size becomes $3L\times 1$. By following the steps from (\ref{likelihood1}) to (\ref{crb}), the CRB for the near-field scenario can be derived, wherein  the derivatives of the steering vector element in (\ref{nf_sv}), i.e., $[\mathbf{a}_\mathrm{NF}']_i = e^{j(\pi  \frac{f_m}{f_c} (i-1)(\vartheta_l   - \frac{ c_0(i-1)\cos^2\tilde{\vartheta}_l}{4 f_c r})   ) }= e^{j\pi (i-1)\theta_{r,m,i}} = e^{j\pi (i-1)(\Delta_{r,l} + \sin\tilde{\vartheta}_l)} $ with respect to $\tilde{\vartheta}_l$, $r_l$ and near-field beam-split $\Delta_{r,l}$ are
	\begin{align}
	\frac{\partial [\mathbf{a}_\mathrm{NF}']_i}{\partial \tilde{\vartheta}_l } &= {-j \pi (i-1) \frac{f_m}{f_c}} (\cos\tilde{\vartheta}_l  \nonumber \\
	& \hspace{20pt} + \frac{ c_0(i-1)\sin 2\tilde{\vartheta}_l}{4 f_c r}) [\mathbf{a}_\mathrm{NF}']_i, \\
	\frac{\partial [\mathbf{a}_\mathrm{NF}']_i}{\partial r_l } &= { \frac{j\pi f_m c_0 (i-1)^2 \cos^2 \tilde{\vartheta}_l}{4 f_c^2 r^2}}  [\mathbf{a}_\mathrm{NF}']_i, \\
	\frac{\partial [\mathbf{a}_\mathrm{NF}']_i}{\partial \Delta_{r,l}} &= {j \pi (i - 1 )} [\mathbf{a}_\mathrm{NF}']_i.
	\end{align}

	%
	%
	%
	%
	%
	%
	%

	%

	\bibliographystyle{IEEEtran}
	\bibliography{IEEEabrv,references_104}
	
	\begin{IEEEbiography}[{\includegraphics[width=1in,height=1.25in,clip,keepaspectratio]{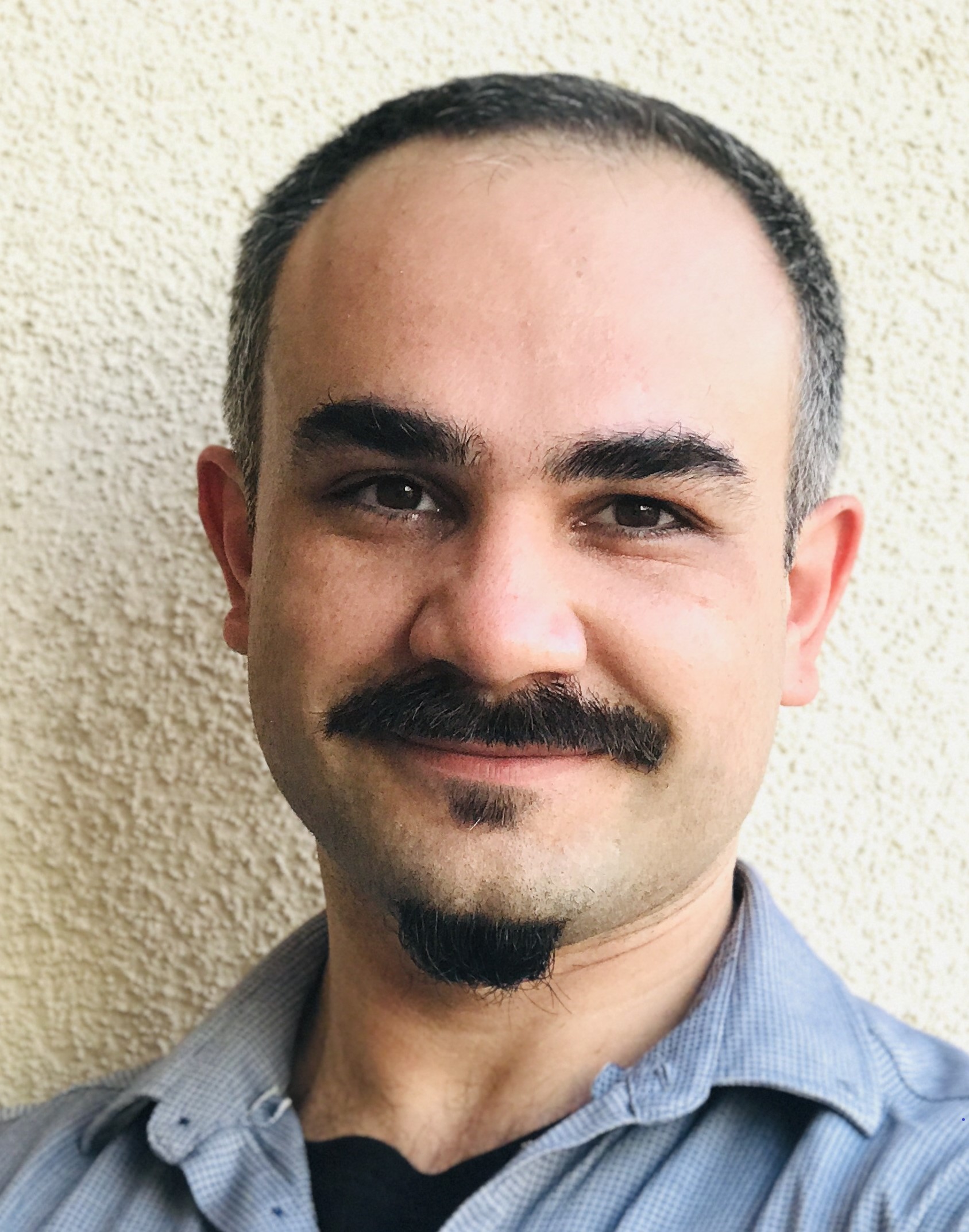}}]{Ahmet M. Elbir} (Senior Member, IEEE) received the B.S. degree with Honors from Firat University in 2009 and the Ph.D. degree from Middle East Technical University (METU) in 2016, both in electrical engineering. Hes held Visiting Postdoctoral Researcher  positions at Koc University, Istanbul, Turkey in 2020-2021, and Carleton University, Canada in 2022-2023. He has been a senior researcher at Duzce University, Turkey in 2016-2022. Currently, he is a Research Fellow at University of Luxembourg, Luxembourg, and University of Hertfordshire, U. K. He is the recipient of 2016 METU best Ph.D. thesis award for his doctoral studies, and the IET Radar, Sonar \& Navigation Best Paper Award in 2022. He serves as an Associate Editor for IEEE Access since 2018, and the Lead Guest Editor for the \textsc{IEEE Journal of Selected Topics on Signal Processing} and \textsc{IEEE Wireless Communications} on Near-field Communications and Signal Processing. His research interests include array signal processing, sparsity-driven convex optimization, signal processing for communications and deep learning for array signal processing.
	\end{IEEEbiography}
	\begin{IEEEbiography}[{\includegraphics[width=1in,height=1.25in,clip,keepaspectratio]{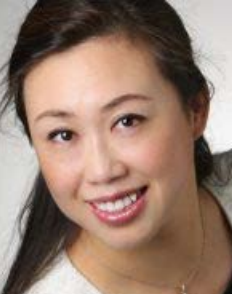}}]{Wei Shi} (Member, IEEE) received the Bachelor 	of Computer Engineering degree from the Harbin 	Institute of Technology (HIT) and the master's and Ph.D. degrees in computer science from Carleton University, Ottawa, ON, Canada. She is currently an Associate Professor with the School of	Information Technology, cross appointed to the
		Department of Systems and Computer Engineering, Faculty of Engineering and Design, Carleton 	University. She is specialized in the design and 	analysis of fault tolerance algorithms addressing security issues in distributed	environments, such as data-center networks, clouds, mobile agents and actuator systems, wireless sensor networks, as well as critical infrastructures,	such as power grids and smart cities. She has also been conducting research	in data privacy and big data analytics. She is also a Professional Engineer	licensed in Canada.
	\end{IEEEbiography}

	\begin{IEEEbiography}[{\includegraphics[width=1in,height=1.25in,clip,keepaspectratio]{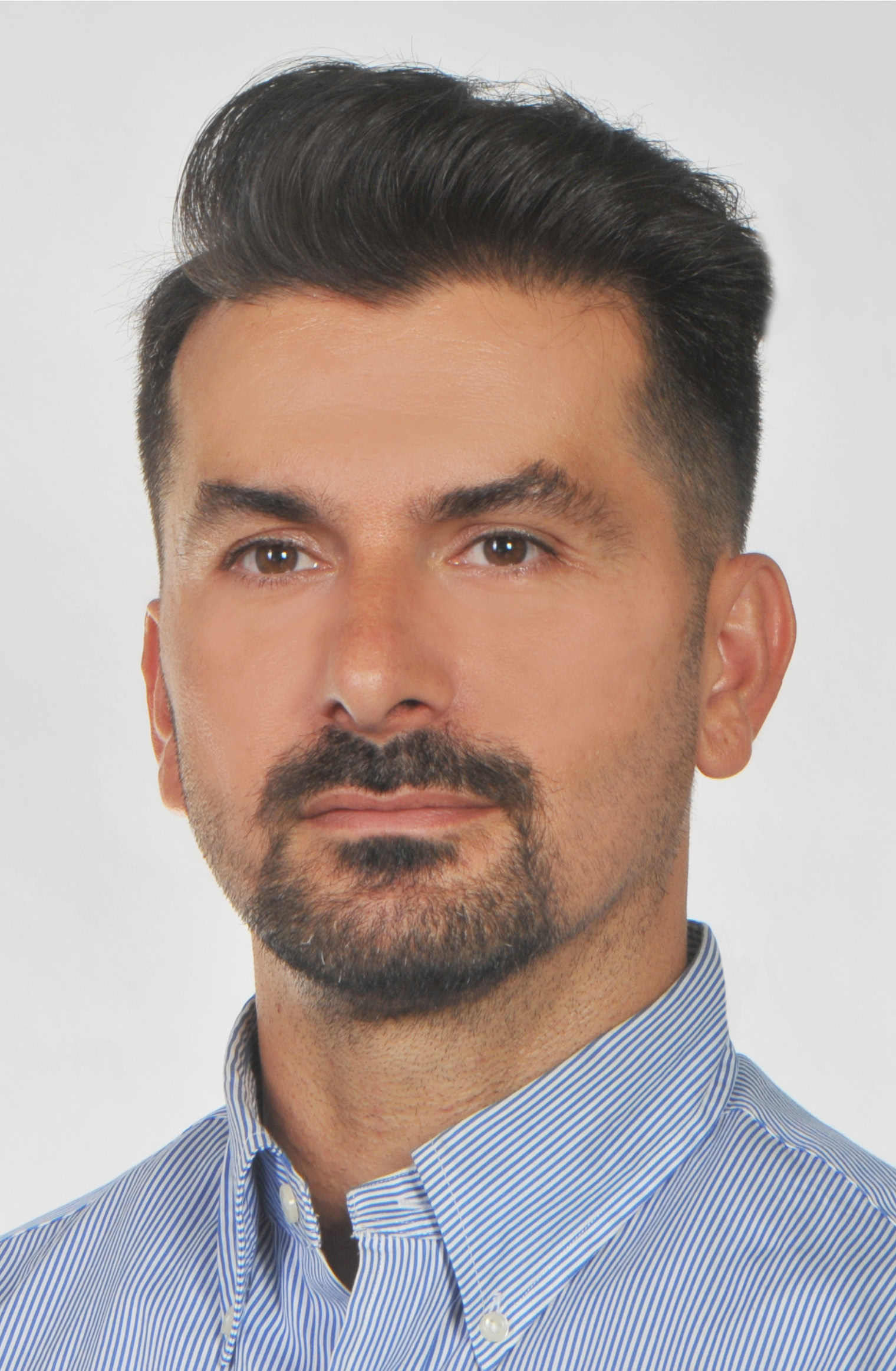}}]{Anastasios K. Papazafeiropoulos} (Senior Member, IEEE) received the B.Sc. degree (Hons.) in physics, the M.Sc. degree (Hons.) in electronics and computers science, and the Ph.D. degree from the University of Patras, Greece, in 2003, 2005, and 2010, respectively. From 2011 to 2012 and from 2016 to 2017, he was a Postdoctoral Research Fellow with the Institute for Digital Communications, University of Edinburgh, Edinburgh, U.K. From 2012 to 2014, he was a Research Fellow with Imperial College London, London, U.K. He has been involved in several EPSRC and EU FP7 projects, including HIATUS and HARP. His research interests include machine learning for wireless communications, intelligent reflecting surfaces, massive MIMO, heterogeneous networks, 5G and beyond wireless networks, full-duplex radio, mm-wave communications, random matrix theory, hardware-constrained communications, and performance analysis of fading channels. He is currently a Vice-Chancellor Fellow with the University of Hertfordshire, Hertfordshire, U.K. He is also a Visiting Research Fellow with the SnT University of Luxembourg, Luxembourg. He was the recipient of a Marie Curie fellowship (IEF-IAWICOM).
	\end{IEEEbiography}

	\begin{IEEEbiography}[{\includegraphics[width=1in,height=1.25in,clip,keepaspectratio]{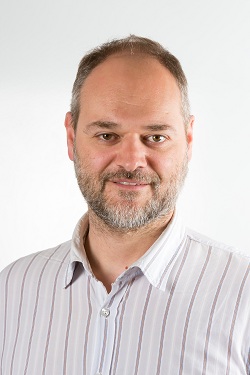}}]{Pandelis Kourtessis} (Member, IEEE)  is currently the Director of the Centre for Engineering Research and Reader in Communication Networks, University of Hertfordshire, U.K., leading the activities of the Networks Engineering Research Group into Communications and Information Engineering, including next generation passive optical networks, optical and wireless MAC protocols, 5G RANs, software-defined network \& network virtualization 5G and satellite networks and more recently machine learning for next generation networks. His funding ID includes EU COST, FP7, H2020, European Space Agency (ESA), UKRI, and industrially funded projects. He has published more than 80 papers at peer-reviewed journals, peer-reviewed conference proceedings, and international conferences. His research has received coverage at scientific journals, magazines, white papers, and international workshops. He has served as the general chair, a co-chair, a technical program committee member, and at the scientific committees and expert groups for IEEE workshops and conferences, European technology platforms and European networks of excellence. He has been the co-editor of a Springer book and a chapter editor of an IET book on softwarization for 5G.
	\end{IEEEbiography}

	\begin{IEEEbiography}[{\includegraphics[width=1in,height=1.25in,clip,keepaspectratio]{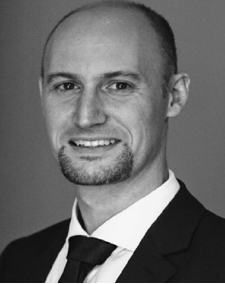}}]{Symeon Chatzinotas}  (Fellow, IEEE)  received the M.Eng. degree in
		telecommunications from the Aristotle University
		of Thessaloniki, Thessaloniki, Greece, in
		2003, and the M.Sc. and Ph.D. degrees in
		electronic engineering from the University of
		Surrey, Guildford, U.K., in 2006 and 2009,
		respectively. He is currently a Full-Professor,
		and the Deputy Head of the SIGCOM Research
		Group, Interdisciplinary Centre for Security,
		Reliability, and Trust, University of Luxembourg,
		Esch-sur-Alzette, Luxembourg, and a Visiting Professor with the University
		of Parma, Parma, Italy. His research interests include multiuser information
		theory, cooperative/ cognitive communications, and wireless network
		optimization. He has been involved in numerous research and development
		projects with the Institute of Informatics Telecommunications, National
		Center for Scientific Research Demokritos, Institute of Telematics and
		Informatics, Center of Research and Technology Hellas, and Mobile
		Communications Research Group, Center of Communication Systems
		Research, University of Surrey. He has coauthored more than 400 technical
		papers in refereed international journals, conferences and scientific books.
		He was the co-recipient of the 2014 IEEE Distinguished Contributions to
		Satellite Communications Award, the CROWNCOM 2015 Best Paper Award,
		and the 2018 EURASIP JWCN Best Paper Award. He is currently on the
		Editorial Board of the IEEE OPEN JOURNAL OF VEHICULAR TECHNOLOGY
		and the International Journal of Satellite Communications and Networking.
	\end{IEEEbiography}
	
\end{document}